\documentclass[twocolumn,showpacs,preprintnumbers,superscriptaddress,prb,aps,psfig]{revtex4}

\usepackage{graphicx}
\usepackage{epstopdf}
\usepackage{float}
\usepackage{color}
\usepackage{amsmath}

\begin{document}

\title{Antiferromagnetic spin correlations and pseudogap-like behavior in Ca(Fe$_{1-x}$Co$_x$)$_2$As$_2$ studied by  $^{75}$As nuclear magnetic resonance and anisotropic resistivity}

\author{J.~Cui}
\affiliation{Ames Laboratory, U.S. DOE, Ames, Iowa 50011, USA}
\affiliation{Department of Chemistry, Iowa State University, Ames, Iowa 50011, USA}
\author{B. Roy}
\affiliation{Ames Laboratory, U.S. DOE, Ames, Iowa 50011, USA}
\affiliation{Department of Physics and Astronomy, Iowa State University, Ames, Iowa 50011, USA}
\author{M.~A.~Tanatar}
\affiliation{Ames Laboratory, U.S. DOE, Ames, Iowa 50011, USA}
\affiliation{Department of Physics and Astronomy, Iowa State University, Ames, Iowa 50011, USA}
\author{S.~Ran}
\affiliation{Ames Laboratory, U.S. DOE, Ames, Iowa 50011, USA}
\affiliation{Department of Physics and Astronomy, Iowa State University, Ames, Iowa 50011, USA}
\author{S.~L.~Bud'ko}
\affiliation{Ames Laboratory, U.S. DOE, Ames, Iowa 50011, USA}
\affiliation{Department of Physics and Astronomy, Iowa State University, Ames, Iowa 50011, USA}
\author{R.~Prozorov}
\affiliation{Ames Laboratory, U.S. DOE, Ames, Iowa 50011, USA}
\affiliation{Department of Physics and Astronomy, Iowa State University, Ames, Iowa 50011, USA}
\author{P.~C.~Canfield}
\affiliation{Ames Laboratory, U.S. DOE, Ames, Iowa 50011, USA}
\affiliation{Department of Physics and Astronomy, Iowa State University, Ames, Iowa 50011, USA}
\author{Y.~Furukawa}
\affiliation{Ames Laboratory, U.S. DOE, Ames, Iowa 50011, USA}
\affiliation{Department of Physics and Astronomy, Iowa State University, Ames, Iowa 50011, USA}

\date{\today}

\begin{abstract} 
      We report $^{75}$As nuclear magnetic resonance (NMR) measurements of single-crystalline Ca(Fe$_{1-x}$Co$_x$)$_2$As$_2$ ($x$  = 0.023, 0.028, 0.033, and 0.059) annealed at 350~$^{\circ}$C for 7 days.
      From the observation of a characteristic shape of $^{75}$As NMR spectra in the stripe-type antiferromagnetic (AFM) state, as in the case of $x$ = 0 ($T_{\rm N}$ = 170 K), clear evidence for the commensurate AFM phase transition with the concomitant structural phase transition  is observed in $x$ = 0.023 ($T_{\rm N}$ = 106 K) and $x$ = 0.028 ($T_{\rm N}$ = 53 K).  
      Through the temperature dependence of the Knight shifts and the nuclear spin lattice relaxation rates (1/$T_1$), although stripe-type AFM spin fluctuations are realized in the paramagnetic state as in the case of other iron pnictide superconductors, we found a gradual decrease of the AFM spin fluctuations below a crossover temperature $T^*$ which was nearly independent of  Co-substitution concentration, and is attributed to a pseudogap-like behavior in the spin excitation spectra of these systems. 
   The $T^*$ feature finds correlation with features in the temperature-dependent inter-plane resistivity, $\rho_c(T)$, but not with the in-plane resistivity $\rho _a (T)$. 
    The temperature evolution of anisotropic stripe-type AFM spin fluctuations are tracked in the paramagnetic and pseudogap phases by the 1/$T_1$ data measured under magnetic fields parallel and perpendicular to the $c$ axis.
       Based on our NMR data, we have added a pseudogap-like phase to the magnetic and electronic phase diagram of Ca(Fe$_{1-x}$Co$_x$)$_2$As$_2$.

\end{abstract}

\pacs{74.70.Xa, 76.60.-k, 75.50Ee, 74.62.Dh}
\maketitle

   \section{Introduction} 

   After the discovery of superconductivity in substituted transition metal pnictides, much attention has been paid to  understanding of the interplay between magnetism and superconductivity in these new materials.\cite{Kamihara2008, Johnston2010, Canfield2010, Stewart2011}  
    Among the iron pnictide superconductors, $A$Fe$_2$As$_2$ ($A$ = Ca, Ba, and Sr), known as "122" compounds with a ThCr$_2$Si$_2$-type structure at room temperature, has been one of the most widely studied systems in recent years.\cite{Johnston2010,Canfield2009, Canfield2010,Stewart2011,Ni2008_1,Ni2008}   
    Application of pressure and carrier doping are considered to play an important role in the suppression of the antiferromagnetic (AFM) ordering and the appearance of the high temperature superconducting (SC) phase.
    These tuning parameters produce the well-known phase diagram of the Fe-based superconductors: an AFM ordering temperature $T_{\rm N}$  is suppressed continuously with doping or pressure application, and  an SC state emerges with the transition temperature $T_{\rm c}$  varying as a function of the tuning parameters.\cite{Johnston2010,Canfield2010,Stewart2011,Ni2008_1,Ni2008}

    Among the 122 compounds,  CaFe$_2$As$_2$ is known to be extremely sensitive to an application of pressure and is considered to be a system with strong coupling of the magnetic and structural phase transitions exhibiting an AFM ordering of the Fe moments at $T_{\rm N}$ = 170 K  with a concomitant structural phase transition to a low temperature orthorhombic phase. \cite{Ni2008, Goldman2008, Canfield2009}   
    Under ambient pressure, substitutions of Fe by Co, Ni and others induce superconductivity in CaFe$_2$As$_2$ with $T_{\rm c}$  up to $\sim$ 15 K. \cite{Canfield2009,Kumar2009_1,Kumar2009_2,Ran2012}   
     Under a pressure of just a few kilobars, the orthorhombic AFM  phase was replaced by a non-magnetic, collapsed tetragonal  phase.\cite{Canfield2009,Torilachvili2008,Lee,Yu}
    The collapsed tetragonal phase in CaFe$_2$As$_2$ is characterized by a $\sim$10 $\%$ reduction in the tetragonal $c$ lattice constant, from the value in the high temperature tetragonal  phase, along with the absence of  AFM ordering.\cite{Kreyssig2008,Goldman2009_2, Ran2011}

\begin{figure}[b]
\includegraphics[width=8.5cm]{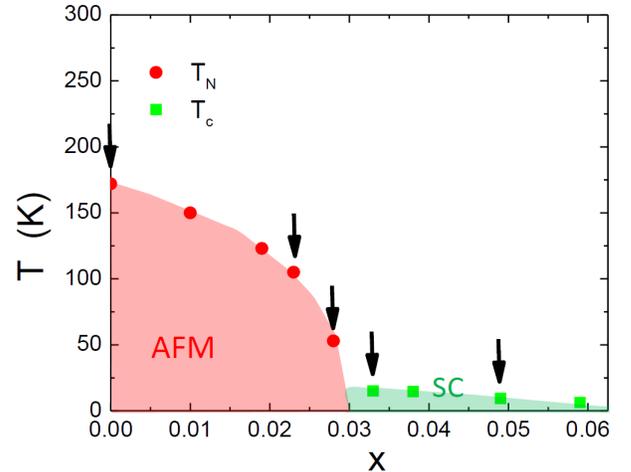} 
\caption{(Color online) Phase diagram of Ca(Fe$_{1-x}$Co$_x$)$_2$As$_2$ in the case of the crystals annealed at $T_{\rm a}$ = 350~$^{\circ}$C for 7 days and then quenched.\cite{Ran2012} 
$T_{\rm N}$ and $T_{\rm c}$ are from Ref. \onlinecite{Ran2012}.
AFM and SC represent the antiferromagnetic ordered state and superconducting phase.  
   Arrows indicate the Co substituted samples used in the present work.}
\label{fig:As-spectrum}
\end{figure}

    Recently it was shown that, by careful combination of Co substitution and post growth annealing and quenching, Ca(Fe$_{1-x}$Co$_x$)$_2$As$_2$ can be systematically tuned to have one of four different ground states: orthorhombic AFM, superconducting, tetragonal paramagnetic  and collapsed tetragonal states.\cite{Ran2012,Ran2011}
      Figure 1 shows the typical phase diagram of Ca(Fe$_{1-x}$Co$_x$)$_2$As$_2$ in the case of the crystals annealed at $T_{\rm a}$ = 350~$^{\circ}$C for 7 days and then quenched.\cite{Ran2012}
      With Co substitution, the AFM state with $T_{\rm N}$ = 170 K at $x$ = 0 is suppressed to 53 K at $x = 0.028$ and then an SC phase shows up with a highest $T_{\rm c}$ $\sim$ 15 K at $x = 0.033$.
      Although the phase diagram is, in some ways similar to the case of Ba(Fe$_{1-x}$Co$_x$)$_2$As$_2$,  Ca(Fe$_{1-x}$Co$_x$)$_2$As$_2$ shows a coincident, first order, structural and magnetic phase transition at the same temperature and does not show  any splitting of the phase transitions upon Co substitution, and no coexistence of the AFM and SC has been reported, \cite{Ran2012}
   while Ba(Fe$_{1-x}$Co$_x$)$_2$As$_2$ system exhibits a clear splitting of those transition lines, and the coexistence of AFM are SC has been found. \cite{Ni2008,Chu2009}

      Nuclear magnetic resonance (NMR) has been known to be a microscopic probe suitable to investigate static spin susceptibility, magnetic order and low energy spin excitations for Fe pnictides superconductors.\cite{Johnston2010,Ishida2009,Ma2013} 
     The NMR spectrum gives us information on static magnetic properties through the hyperfine interactions of the nuclei with Fe spins while the nuclear spin lattice relaxation  rate (1/$T_1$) is related to the power spectrum of the hyperfine field fluctuations produced by the Fe spins.\cite{Johnston2010,Ishida2009,Ma2013} 
     Previous $^{75}$As NMR studies of the parent material CaFe$_2$As$_2$ showed clear splittings of $^{75}$As NMR lines due to a hyperfine field produced by Fe moments below N\'eel temperature $T_{\rm N}$ = 170 K, demonstrating a phase transition from a high temperature paramagnetic state to a low temperature  stripe-type AFM state.\cite{BaekCaFe2As2_P, BaekCaFe2As2,Curro2009CaFe2As2}
     Suppression of the Fe spin correlations in the collapsed tetragonal phase in CaFe$_2$As$_2$ was also revealed by $^{75}$As NMR.\cite{Kawasaki2010,Furukawa2014}

      In the case of Co substituted CaFe$_2$As$_2$, Baek ${\it et~al.}$ reported nuclear spin lattice relaxation rates (1/$T_1$) of $^{75}$As nuclear quadrupole resonance (NQR) as a function of temperature in Ca(Fe$_{1-x}$Co$_x$)$_2$As$_2$ grown out of Sn flux, showing a gradual decrease of 1/$T_1T$  below a crossover temperature ($T^*$) in the under- and optimally-substituted regions.\cite{Baek2011}  
   The decrease in 1/$T_1T$ has been attributed to pseudogap-like phenomenon and the crossover temperature $T^*$  shows a strong substitution dependence, falling to zero near optimum substitution. 
     Pseudogap-like behavior has been reported in the isostructural Co substituted BaFe$_2$As$_2$ from temperature dependence of Knight shift and 1/$T_1T$ of $^{75}$As NMR measurements\cite{Ning2010}  which provide important information about static and dynamical magnetic properties, in addition to NQR measurements. 
   Furthermore, NMR measurements, in particular 1/$T_1$ measurements under different magnetic field directions,  provide more detailed information about magnetic fluctuations.\cite{Ishida2009}
     Thus, using NMR techniques, detailed studies of Co substitution effects on static and dynamical magnetic properties in CaFe$_2$As$_2$ are important and of a great deal of interest.  
       However, no systematic NMR data on Co substituted  CaFe$_2$As$_2$ have been reported up to now. 
 
   It was found previously,\cite{Tanatar2010} that the temperature-dependent NMR Knight shift in substituted Ba122 compounds shows a correlation with the temperature-dependent resistivity, particularly for the inter-plane transport direction, $\rho _c(T)$. 
    This correlation was interpreted as evidence for magnetic character of scattering in the compounds \cite{Rudoping} and indication of a partial charge gap (pseudogap) \cite{Tanatar2010,pseudogap2,BaKinterplane} developing at high temperatures. 
    The pseudogap-like behavior has been also reported in iron pnictides by other experimental techniques such as angle-resolved photoemission spectroscopy and optical measurements.\cite{Xu2011,Shimojima2014,Moon2012}
    Change of alkali earth element in the 122 family from Ba to Ca, leading to inevitable change of the Fermi surface, \cite{Liu2011,Dhaka2014} can bring additional insight into the origin of pseudogap and its dependence on material properties.

      In this paper,  we report a comprehensive study of the $^{75}$As NMR in Ca(Fe$_{1-x}$Co$_x$)$_2$As$_2$  and its comparison with inter-plane transport properties.
     Here we used single crystals grown out of a FeAs/CoAs flux since the effects of Co substitution on the crystals grown out of Sn flux have issues with solubility, reproducibility, and inhomogeneity, \cite{Harnagea2011,Matusiak2010,Hu2012}  while one can minimized these problems in Co substituted CaFe$_2$As$_2$ grown out of an FeAs/CoAs flux by systematically control annealing/quenching temperatures.\cite{Ran2011,Ran2012}
     We present the temperature and the $x$ dependence of NMR spectra from which we derive information about the hyperfine and quadrupole interactions at the $^{75}$As sites exhibiting microscopic evidence of a simultaneous stripe-type AFM and structural phase transition in Co substituted CaFe$_2$As$_2$. 
    We also report the temperature and $x$ dependence of nuclear relaxation rates that provide the pseudogap-like phase in the phase diagram as shown in Fig. 12 where the crossover temperature $T^*$ is found to be nearly independent of $x$, in contrast to the previous report.\cite{Baek2011} 
    We support this interpretation by observation of features in the temperature-dependent inter-plane transport.

 \section{Experimental}
    
   The single crystals of Ca(Fe$_{1-x}$Co$_x$)$_2$As$_2$  ($x$ = 0.023, 0.028, 0.033 and 0.059) used in this study were grown out of a FeAs/CoAs flux,\cite{Ran2011,Ran2012} using conventional high temperature growth techniques.\cite{Canfield_book, Canfield_1992}
   Subsequent to growth, the single crystals were annealed at $T_{\rm a}$ =  350~$^{\circ}$C for 7 days and then quenched.  
   For $x$ = 0, we used the single crystal annealed at $T_{\rm a}$ = 400~$^{\circ}$C for 24 hours. 
   The Co substitution levels of the single crystals used in this study were determined by a wavelength dispersive x-ray spectroscopy, and the crystals are characterized by magnetic susceptibility,\cite{Ran2012} resistivity\cite{Ran2012} and thermal expansion\cite{Sergey2013} measurements.    
     Details of the growth, annealing and quenching procedures are reported in Refs.~\onlinecite{Ran2012} and \onlinecite{Ran2011}.

    NMR measurements were carried out on $^{75}$As  (\textit{I} = 3/2, $\gamma/2\pi$ = 7.2919 MHz/T, $Q$ =  0.29 Barns)  by using a lab-built, phase-coherent, spin-echo pulse spectrometer.  
   The $^{75}$As-NMR spectra were obtained by sweeping the magnetic field at a fixed frequency $f$ = 53 MHz.
  The magnetic field was applied parallel to either the crystal $c$ axis or the $ab$ plane, and the direction of the magnetic field on the ab plane was not controlled. 
    The origin of the Knight shift, $K$ = 0, of the $^{75}$As nucleus was determined by the $^{75}$As NMR measurements of GaAs. 
    The $^{75}$As 1/$T_{\rm 1}$ was measured with a recovery method using a single $\pi$/2 saturation $rf$ pulse. 
   The $1/T_1$ at each $T$ was determined by fitting the nuclear magnetization $M$ versus time $t$  using the exponential functions $1-M(t)/M(\infty) = 0.1 e^ {-t/T_{1}} +0.9e^ {-6t/T_{1}}$ for $^{75}$As NMR, where $M(t)$ and $M(\infty)$ are the nuclear magnetization at time $t$ after the saturation and the equilibrium nuclear magnetization at $t$ $\rightarrow$ $\infty$, respectively. 
     In the paramagnetic states, the nuclear magnetization recovery curves were well fitted by the function in all Co-substituted crystals within our experimental uncertainty. 
     On the other hand, below $T_{\rm N}$ or $T_{\rm c}$, we observed a slight deviation due to short $T_{\rm 1}$ components with unknown in origin.
     A part of NMR data for the parent compound ($x$ = 0) annealed at 400 $^{\circ}$C has been reported previously.\cite{Furukawa2014}  

\begin{figure}[tb]
\includegraphics[width=8.5cm]{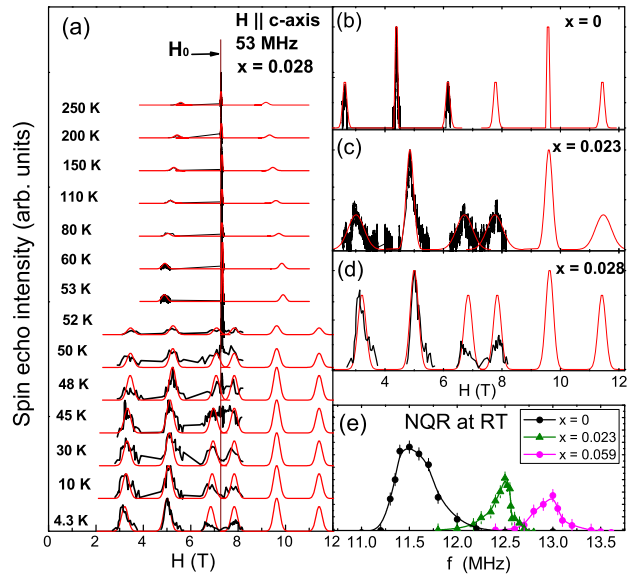} 
\caption{(Color online) (a) Temperature variation of field-swept $^{75}$As NMR spectra for the $x$ = 0.028 Co substituted CaFe$_2$As$_2$ crystal (annealed at $T_{\rm a}$ = 350~$^{\circ}$C for 7 days) at $f$ = 53 MHz for magnetic field parallel to the $c$ axis. 
    The black and red lines are observed and simulated spectra, respectively. 
     Expected lines above 8.5 T are not measured due to the limited maximum magnetic field for our SC magnet.
     Field-swept $^{75}$As NMR spectra in antiferromagnetic state are shown in (b) for $x$ = 0 ($T$ = 4.3 K), (c) $x$ = 0.023 ($T$ = 4.3 K) and (d) $x$ = 0.028 ($T$ = 4.3 K), together with the simulated spectrum for each.
(e) $^{75}$As NQR spectra measured at room temperature. }
\label{fig:As-spectrum}
\end{figure}

      The in-plane resistivity measurements were made in four-probe configuration on samples cut into bars with typical dimensions 1$\times$0.2$\times$0.2 mm$^3$ ($a$$\times$$b$$\times$$c$). 
     Contacts to the samples were made by Sn soldering 50 $\mu$m diameter Ag wires.
Inter-plane resistivity measurements were made in the two-probe sample configuration.\cite{anisotropy} 
     Contacts were covering the whole $ab$ plane area of the $c$ axis samples, typically 0.5$\times$0.5 mm$^2$, while current was flowing along $c$ axis (short dimension typically 0.1 mm). 
    A four-probe scheme was used to measure the resistance down to the contact to the sample, i.e. the sum of the actual sample resistance $R_s$ and contact resistance $R_c$ was measured. 
    These measurements were relying on ultra-low contact resistance on soldered Sn contacts.\cite{SUST,patent}
    Taking into account that $R_s \gg R_c$,  contact resistance represents a minor correction of the order of 1 to 5\%. 
 
     The drawback of the measurement on samples with $a$ (or $b$) $\gg c$ is that any inhomogeneity in the contact resistivity or internal sample connectivity admixes in-plane component due to redistribution of the current. 
    To minimize this effect, we performed measurements of $\rho_c$ on at least 5 samples of each composition. 
     In all cases we obtained qualitatively similar temperature dependences of the electrical resistivity, as represented by the ratio of resistivities at room and low temperatures, $\rho _c (0)/\rho _c (300)$. 
   The resistivity value, however, showed a notable scatter and at room temperature was typically in the range 1 to 2 m$\Omega$cm. 
    For the sake of comparison we selected the samples with the temperature dependence of resistivity least similar to that of $\rho _a(T)$. 
    Typically, these samples had the lowest value of electrical resistivity, as described in detail in Ref.~\onlinecite{anisotropy}. 
   This is important since partial exfoliation increases resistivity values.\cite{anisotropy} 

Because Sn contacts are covering the whole $ab$-plane area of the samples, they can potentially create uncontrolled stress/strain. Due to strong sensitivity to strain, this can lead to non-negligible effect on the features observed in CaFe$_2$As$_2$ compositions. 
For some compositions we performed measurements using Montgomery technique,\cite{Montgomery1,Montgomery2} in which contacts are located at the sample corners, see discussion below. We have not observed any significant effect of the contact-related stress, similar to measurements in parent BaFe$_2$As$_2$ under pressure.\cite{Rudoping}

\section{Results and discussion}
\subsection{$^{75}$As NMR spectra}
     
    In the paramagnetic state of CaFe$_2$As$_2$, the $^{75}$As NMR spectrum exhibits a typical feature of a nuclear spin $I$ = 3/2 with Zeeman and quadrupolar interactions. This results in a sharp central transition and two satellite lines split by the quadrupolar interaction of the As nucleus with the local electric field gradient (EFG).\cite{BaekCaFe2As2,Furukawa2014}
   Just below $T_{\rm N}$, when $H$ is applied parallel to the $c$ axis, each NMR line splits into two lines due to internal field $H_{\rm int}$ (parallel or antiparallel to $H$) which is produced by the Fe spin ordered moments with the stripe-type spin structure. \cite{KitagawaBaFeAs}
  
     In the case of Co substituted crystals with $x$ = 0.023 and 0.028, similar splittings of the NMR lines are observed below $T_{\rm N}$. 
     Figure 2(a)  shows a typical example of temperature evolution of the field-swept $^{75}$As-NMR spectra of the $x$ = 0.028 Co substituted CaFe$_2$As$_2$ crystal for a magnetic field $H$ $\parallel$ $c$ axis. 
    Just below $T_{\rm N}$ = 53 K, the spectra split into the two sets of three lines due to the internal field as in the case of CaFe$_2$As$_2$. 
   The observed spectra are reproduced well by  a simple nuclear spin Hamiltonian 
${\cal H}$ = ${\cal -}$$\gamma$$\hbar$${\vec I}$$\cdot$${\vec H_{\rm eff}}$ +$\frac{h \nu_{\rm Q}}{6}$ [3$I_{\rm z}^{2}$-$I(I+1)$ + $\frac{1}{2}$$\eta$($I_+^2$ +$I_-^2$) ], where $H_{\rm eff}$ is the effective field at the As site (summation of external field $H$ and the internal field $H_{\rm int}$), $h$ is Planck's constant, and $\nu_{\rm Q}$ is nuclear quadrupole frequency defined by $\nu_{\rm{Q}}$ =  $eQV_{ZZ}$/2$h$ where $Q$ is the quadrupole moment of the As nucleus, $V_{ZZ}$ is the EFG at the As site, and $\eta$ is an asymmetric parameter of EFG.\cite{Slichter_book}
   
    From the simulated spectra shown by red lines in Fig. 2(a), we extracted the temperature dependence of $\nu_{\rm Q}$ and $H_{\rm int}$ for $x$ = 0.028, which are shown in Fig. 3 together with the data for $x$ = 0 and 0.023. 
    For $x$ = 0.028, with decreasing temperature, $\nu_{\rm Q}$ increases from 12.2 MHz at $T$ = 250 K to 17.5 MHz at 53 K, shows a sudden jump to 13 MHz just below 53 K and levels off in the antiferromagnetic state at low temperatures. 
    This clearly indicates the first-order structural phase transition at $T_{\rm s}$ = 53 K.  
    In our experiment, we do not observe clear hysteresis within our experimental uncertainty.         
     Similarly $H_{\rm int}$ = 2.25 T at $T$ = 4.3 K decrease slightly with increasing $T$ and then suddenly disappears above 53 K.  
      These results indicate that, even for the Co substitution level of $x$ = 0.028,  the crystal exhibits the AFM ordering at $T_{\rm N}$= 53 K with a concomitant structural phase transition as in the case of CaFe$_2$As$_2$ ($T_{\rm N}$  = 170 K). 
     This is in sharp contrast to the case of Ba(Fe$_{1-x}$Co$_x$)$_2$As$_2$ where the two phase transitions separate ($T_{\rm S}$ $>$  $T_{\rm N}$) upon Co substitution.\cite{Ni2008,Chu2009} 
      These results suggest that couplings between lattice and magnetism in CaFe$_2$As$_2$ are much stronger than in other pnictides such as BaFe$_2$As$_2$.
        $H_{\rm int}$ and $\nu_{\rm Q}$  in a temperature range of $T$ = 50 -- 100 K for $x$ = 0.023 were not measured because  of poor signal intensities.

 \begin{figure}[tb]
 \includegraphics[width=8.0cm]{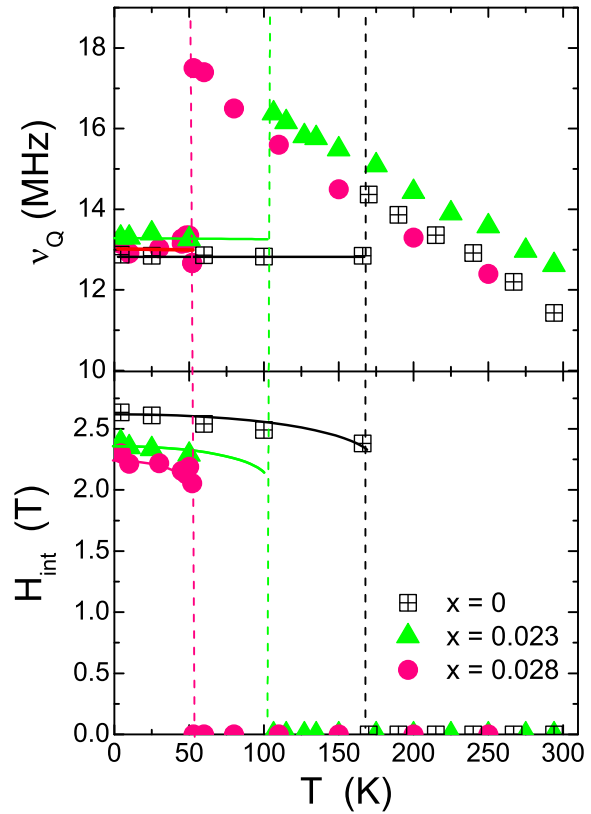} 
 \caption{(Color online) (top) Temperature dependence of quadrupole frequency $\nu_{\rm Q}$ for $x$ = 0.023 and 0.028. 
    (bottom) Temperature dependence of $H_{\rm int}$. 
   The vertical broken lines correspond  to $T_{\rm N}$ = $T_{\rm s}$ for each crystal determined by the magnetic susceptibility measurements.\cite{Ran2012} 
  $H_{\rm int}$ and $\nu_{\rm Q}$  in a temperature range of $T$ = 50 -- 100 K for $x$ = 0.023 were not measured because  of poor signal intensities.  
The solid lines are guides for eyes.  }
 \label{fig:Hint}
 \end{figure}

   Figures 2(b)-(d) show a comparison of $^{75}$As NMR spectra in the stripe-type AFM state for $x$ = 0, 0.023 and 0.028. 
    With Co substitution,  each line broadens but is still well separated, implying that $\nu_{\rm Q}$ and $H_{\rm int}$ can be well defined. 
    This is in contrast to the case of Ba(Fe$_{1-x}$Co$_x$)$_2$As$_2$ where very broad and featureless $^{75}$As NMR lines were observed in antiferromagnetic state for $x$ = 0.02 and 0.04.\cite{Ning2009}
   Clear split lines observed even for $x = 0.028$ indicate that the stripe-type AFM structure is commensurate upon Co substitution in CaFe$_2$As$_2$.
  Similar splitting of $^{75}$As NMR lines in the AFM state in 2 \% Co substituted BaFe$_2$As$_2$ has been observed recently,\cite{Ning2014} consistent with commensurate AFM state.

   Figure 1(e) shows the $^{75}$As NQR spectra in crystals with different Co-substitution levels  at room temperature. 
    The line width ($\sim$ 0.5 MHz) of the spectrum is nearly independent of $x$, which indicates that no significant increase of inhomogeneity in distribution of electronic field gradient (EFG) with Co substitution. 
    In the case of the Ca(Fe$_{1-x}$Co$_x$)$_2$As$_2$ crystals grown with Sn flux, the line width of NQR spectrum increases from 0.4 MHz at $x$ = 0 to 0.95 MHz at $x$ = 0.09 (Ref. \onlinecite{Baek2011}).  
    The smaller line widths indicate a higher degree of homogeneity in crystals grown with FeAs/CoAs flux than that with Sn flux, consistent with the previous report.\cite{Hu2012}

      As shown in Fig. 3, the saturated values of $H_{\rm int}$ decrease slightly from 2.64 $\pm$ 0.05 T at $x$ = 0 to 2.35 $\pm$ 0.1 T for $x$ = 0.023 and to 2.25 $\pm$ 0.1 T for $x$ = 0.028, although $T_{\rm N}$ changes drastically from 170 K for $x$ = 0 to 106 K for $x$ = 0.023 and to 53 K for $x$ = 0.028.
    This is in contrast to the case in Ni substituted Ba(Fe$_{1-x}$Ni$_x$)$_2$As$_2$  where $H_{\rm int}$ = 1.5 T at $x$ = 0  at  As sites decreases upon Ni substitution with a similar reduction of $T_{\rm N}$.\cite{Dioguardi2010}
    The $H_{\rm int}$ at the As sites in the stripe-type AFM state in $A$Fe$_2$As$_2$ ($A$ = Ca, Ba, and Sr)  is known to be expressed as $H_{\rm int}$ = 4 $B_{c}$$\langle$s$\rangle$  where $B_{c}$ is an off-diagonal term in the hyperfine coupling tensor and $\langle$s$\rangle$ the ordered magnetic moments.\cite{KitagawaBaFeAs}
    Using the value $\langle$s$\rangle$ = 0.8 $\mu_{\rm B}$ from neutron scattering measurements\cite{NeutronCaFe2As2} and $H_{\rm int}$ = 2.64 T, $B_{\rm c}$ is estimated to be 0.82 T/$\mu_{\rm B}$/Fe, which is in good agreement with the previously reported value.\cite {Curro2009CaFe2As2,Xiao2012}  
     
    The nearly $x$-independent $H_{\rm int}$ suggests that the ordered Fe magnetic moments are not suppressed drastically, if the hyperfine coupling constant  $B_{c}$ does not change drastically with Co substitution.  
     Assuming that $B_{c}$ is independent of $x$, $\langle$s$\rangle$ is estimated to be 0.72 $\mu_{\rm B}$  and 0.71 $\mu_{\rm B}$ for $x$ = 0.023 and 0.028, respectively. 
    On the other hand, if one assumes that the Fe ordered moments decrease with Co substitution, the nearly $x$-independent $H_{\rm int}$ could be possible only if $B_{c}$ increases drastically Co substitution to compensate for the reduction of $\langle$s$\rangle$.
   However, this is highly unlikely because, as can be seen below,  the $K$-$\chi$ plot analysis reveals a slight change in diagonal terms of the hyperfine coupling tensor with Co substitution, and also 1/$T_1T$, related to the square of hyperfine coupling constants, are nearly $x$ independent at high $T$ suggesting no drastic change in hyperfine coupling constants.
    Thus we consider  that Fe ordered moments are robust with Co substitution.
    The robustness may be explained if $``$intrinsic$"$ $T_{\rm N}$  were nearly independent of $x$.
Here the  $``$intrinsic$"$ $T_{\rm N}$ is a N\'eel temperature in the orthorhombic phase and is considered to be much higher than $T_{\rm S}$ (or $``$observed$"$  $T_{\rm N}$)  as expected from the temperature dependence of $H_{\rm int}$ shown in Fig. 3.
    Since the AFM ordering state can be established in only the orthorhombic phase, the suppression of the $``$observed$"$  $T_{\rm N}$ with $x$ can be due to the reduction of $T_{\rm S}$.  
   Assuming that Co substitution suppresses mainly the structural phase transition temperature but not $``$intrinsic$"$ $T_{\rm N}$, the nearly $x$-independent $H_{\rm int}$ can be expected, and thus the Fe ordered moments can be almost independent of $x$. 
    These results are in sharply contrast to the case of Ba(Fe$_{1-x}$Co$_x$)$_2$As$_2$ where the neutron scattering measurements show a monotonic decrease in Fe ordered moments with Co substitution.\cite{Fernandes2010}
     It will be interesting to perform neutron scattering and/or M\"ossbauer measurements to confirm the robustness of the Fe ordered moments in Ca(Fe$_{1-x}$Co$_x$)$_2$As$_2$.

\begin{figure}[tb]
\includegraphics[width=8.0 cm]{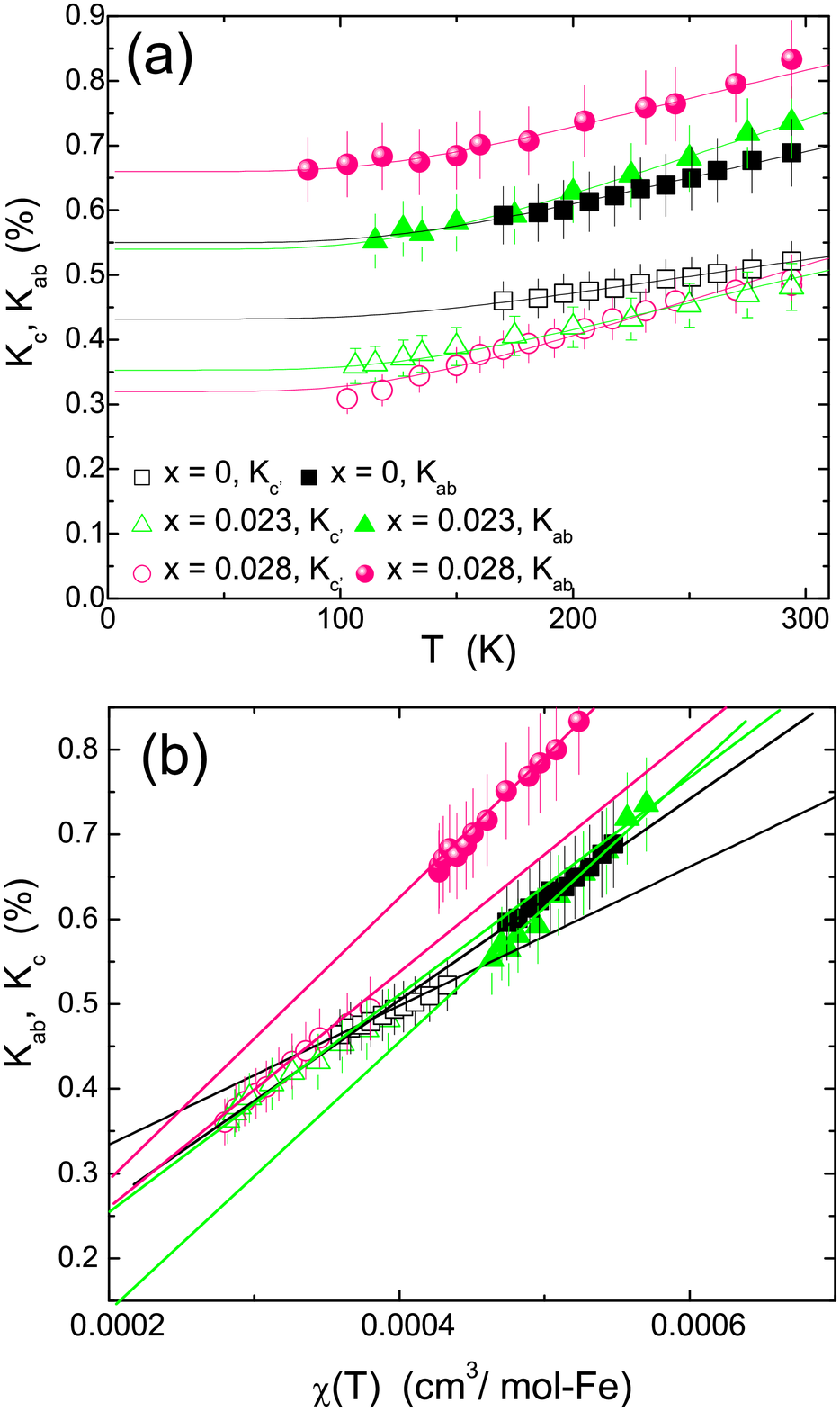} 
\caption{(Color online) (a) Temperature $T$ dependence of $^{75}$As NMR shifts $K_{ab}$ and $K_{c}$ 
for Ca(Fe$_{1-x}$Co$_x$)$_2$As$_2$. 
     The solid lines are fitting results with a thermal activation form $K$ $\sim$ exp(--$\Delta$/$k_{\rm B}T$) with $\Delta$/$k_{\rm B}$  = 510 K for $x$ = 0, and 490 K for $x=0.023$ and 0.028, respectively.  
(b) $K$ versus magnetic susceptibility $\chi(T)$ plots for the corresponding $ab$ and $c$ components of $K$ in Ca(Fe$_{1-x}$Co$_x$)$_2$As$_2$ with $T$ as an implicit parameter. 
  The solid lines are linear fits.
}
\label{fig:T-K}
\end{figure}

   Figure \ \ref{fig:T-K}(a)  shows the $x$ and $T$ dependence of the Knight shift, $K_{\rm ab}$ for $H$  parallel to the $ab$ plane and $K_{\rm c}$ for $H$  perpendicular to the $c$ axis, respectively, where the second order quadrupole shift was corrected  in  $K_{\rm ab}$.\cite{Slichter_book, Furukawa2014}
   With decreasing $T$, all Knight shifts decrease down to $T_{\rm N}$ for each crystal,  similar to $\chi(T)$ data shown in Ref. \onlinecite{Ran2012} for these samples. 
     It is noted that $K_c$ $\sim$ 0.3 -- 0.5 $\%$ for $x$ = 0, 0.023 and 0.028 is greater than $K_c$ = 0.2 -- 0.3 $\%$ for Sn-flux  CaFe$_2$As$_2$.\cite{BaekCaFe2As2} 
   The possible small misalignment of the crystal orientation, the deviation of $H$ from $H$ $\parallel$ $c$ axis or $ab$ plane, will results in additional corrections in second order quadrupole shifts for the central line position of $^{75}$As NMR spectrum, which produces a small change in  the absolute value of the $K$. 
   Although we tried to set the crystal $H$ $\parallel$ $c$ or $H$ $\parallel$ $ab$ as precise as possible, a small misalignment of the crystal orientation is still possible.  
%    Since a small deviation of $H$ from $H$ $\parallel$ $c$ axis or $H$ $\parallel$ $ab$ plane produces additional corrections in second order quadrupole shifts for the central line position of $^{75}$As NMR spectrum, the absolute value of the $K$ would be modified.  
   Since the temperature dependence of $K$ will not be affected much, we focus on mainly on the temperature dependence of $K$'s exhibiting the gradual decrease upon cooling.
     Similar temperature dependence of  Knight shifts (or macroscopic magnetic susceptibility) were reported previously for various Fe based superconductors such as Ba(Fe$_{1-x}$Co$_x$)$_2$As$_2$ (Ref.~\onlinecite{Ning2010}), LaFeAsO$_{\rm 1-x}$F$_x$ (Refs. \onlinecite{Klingeler2010} and \onlinecite{Grafe2008}) and FeSe (Ref. \onlinecite{Kotegawa2008}).
   The gradual decreases in $K$  indicate gradual suppressions of the $q$ = 0 component of the spin susceptibility on cooling and  were fitted by a phenomenological thermal activation form $K \propto $ exp$(-\Delta/k_{\rm B}T$).
    Using the equation we estimate $\Delta/k_{\rm B}$ = 490$-$510 K which is almost independent of $x$ in the Co substituted compounds. 
       The solid lines in the Fig. 4 are fitting results. 
   This value is comparable to the previous estimates of 450 K in Ba(Fe$_{1-x}$Co$_x$)$_2$As$_2$ (Ref. \onlinecite{Ning2010})  and 435 K in K$_x$Fe$_{2-x}$Se$_2$ (Ref. \onlinecite{Torchetti2011}). 
    We also tentatively fitted the data for $x=0$ from Ref. \onlinecite{BaekCaFe2As2} with the formula which produces a relatively large value of $\Delta \sim$ 650 K.

 \begin{figure}[tb]
 \includegraphics[width=8.0cm]{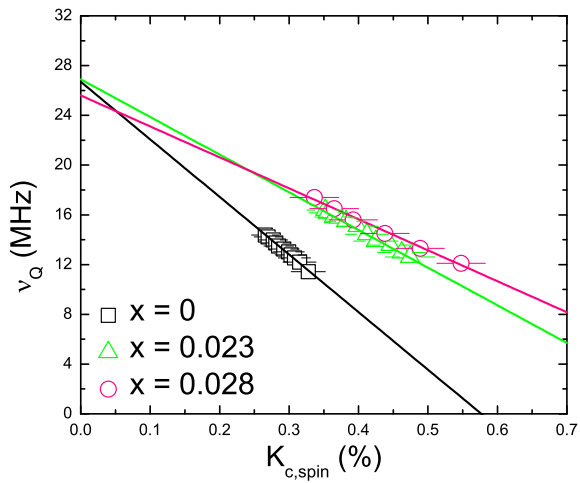} 
 \caption{(Color online) $\nu_{\rm Q}$ versus $K_{c, spin}$ plots  for  $x$ = 0 (black), $x$ = 0.023 (light green) and  $x$ = 0.028 (pink) in Ca(Fe$_{\rm 1-x}$Co$_{x}$)$_2$As$_2$ with $T$ as an implicit parameter. 
  The solid lines are linear fits.
}
 \label{fig:K-nQ}
 \end{figure}

     The diagonal terms of the hyperfine coupling tensor $A_{\rm{hf}}$ can be estimated by $K$-$\chi$ plot analysis.
     Since the spin part of $K$, $K_{\rm spin}$, is proportional to the spin susceptibility $\chi_{\rm spin}$ through the diagonal term of the hyperfine coupling tensor $A_{\rm{hf}}$ giving $K_{\rm spin}$ =  $\frac{A_{\rm {hf}}}{N_{\rm A}}\chi_{\text{spin}}(T)$, where $N_{\rm A}$ is Avogadro's number,  the slope of the $K$-$\chi$ plot gives an estimate of  $A_{\rm{hf}}$.
     Figure \ \ref{fig:T-K}(b)  plots $K_{ab}$ and $K_{c}$ against the corresponding $\chi_{ab}$ and $\chi_{c}$, respectively, for each sample with $T$ as an implicit parameter. 
     All $K_{ab}$ and $K_c$ are seen to vary linearly with the corresponding $\chi$ and the hyperfine coupling constants are estimated to be $A_{c}= (-12.2 \pm 2.0) $ kOe/$\mu_{\rm B}$/Fe, ($-14.6 \pm 1.4)$  kOe/$\mu_{\rm B}$/Fe  and ($-15.7\pm 1.4)$  kOe/$\mu_{\rm B}$/Fe and $A_{ab} = (-17.9 \pm 2.2) $ kOe/$\mu_{\rm B}$, ($-19.0 \pm 2.0)$  kOe/$\mu_{\rm B}$/Fe and ($-20.5\pm 3.0)$  kOe/$\mu_{\rm B}$/Fe, for $x$ = 0, 0.023 and 0.028, respectively.
    One does not observe a significant change in the diagonal term of the hyperfine coupling tensor in the Co substituted compounds within our experimental uncertainty,  although both $A_{c}$ and $A_{ab}$ seems to be increased slightly with the Co substitution.

 \begin{figure*}[tb]
 \includegraphics[width=18.0cm]{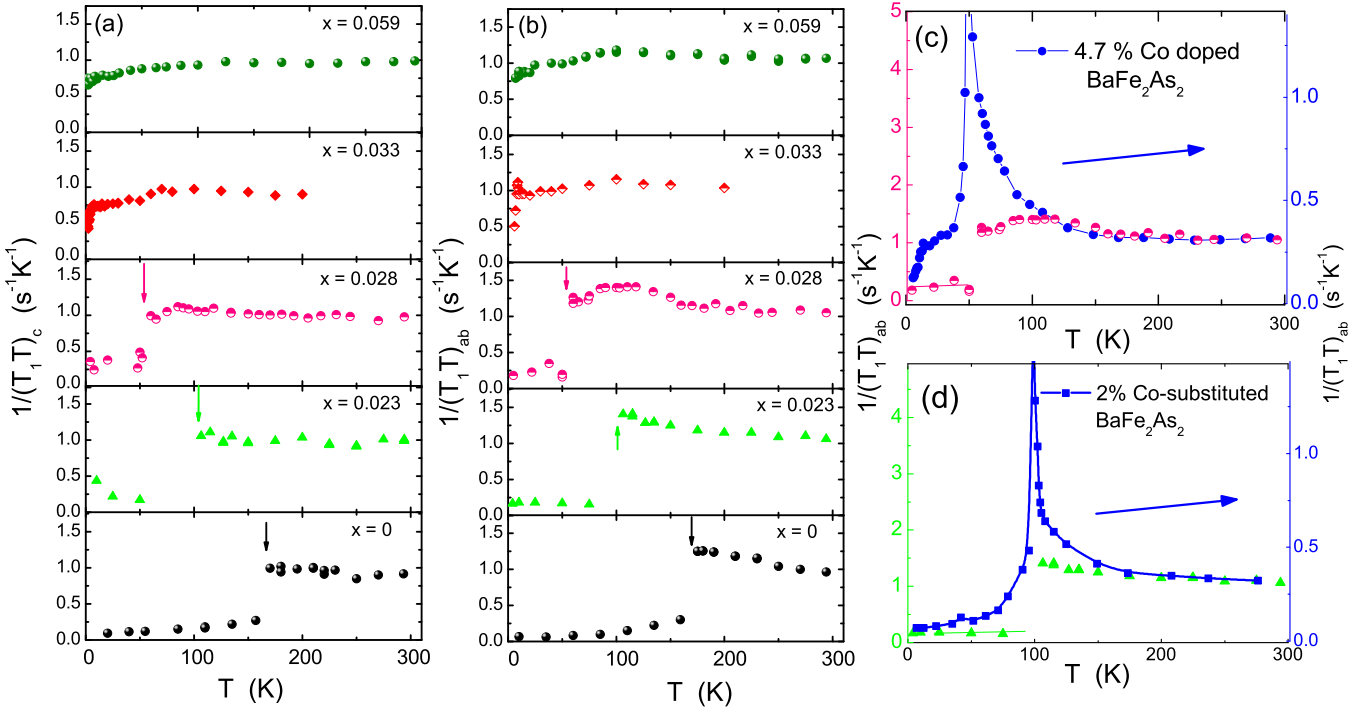} 
 \caption{(Color online) Temperature dependence of 1/$T_1T$ in Ca(Fe$_{\rm 1-x}$Co$_{x}$)$_2$As$_2$.
 (a) $H$ $\parallel$ $c$ axis. (b) $H$ $\parallel$ $ab$ plane. 
  The arrows indicate  $T_{\rm N}$ for $x$ = 0 (black), $x$ = 0.023 (light green) and  $x$ = 0.028 (pink) determined by the magnetic susceptibility measurements.\cite{Ran2012} 
In (c) and (d),  we compare the temperature dependence of (1/$T_1T$)$_{ab}$ for $H$ $\parallel$ $ab$ plane in Ca(Fe$_{\rm 1-x}$Co$_{x}$)$_2$As$_2$ with that of (1/$T_1T$)$_{ab}$ in Ba(Fe$_{\rm 1-x}$Co$_{x}$)$_2$As$_2$.
(c) $x$ = 0.047 (blue circles) in Ba(Fe$_{\rm 1-x}$Co$_{x}$)$_2$As$_2$ with $T_{\rm N}$ = 50 K  and $T_{\rm C}$ = 15 K  (data from Ref. \onlinecite{Wiecki2015}), together with $x$ = 0.028 (pink symbols) in Ca(Fe$_{\rm 1-x}$Co$_{x}$)$_2$As$_2$ with $T_{\rm N}$ = 53 K.  
(d) $x$ = 0.02 (blue squares) in Ba(Fe$_{\rm 1-x}$Co$_{x}$)$_2$As$_2$ with $T_{\rm N}$ = 99 K (data from Ref. \onlinecite{Ning2014}), together with $x$ = 0.023 (green triangles) in Ca(Fe$_{\rm 1-x}$Co$_{x}$)$_2$As$_2$ with $T_{\rm N}$ = 106 K. }
 \label{fig:T1T}
 \end{figure*}

       As shown in Fig.\ \ref{fig:K-nQ}, we have observed that the $^{75}$As quadrupole frequency, $\nu_{\rm Q}$ varies linearly with the spin part of Knight shift in the paramagnetic phase with the relation $\nu_Q$ = $\nu_{Q0}$ + $\alpha$$K_{\rm spin}$. 
      As can be seen, $\alpha$  decreases from $-$46 MHz/$\%$  for $x$ = 0 to $-$30 MHz/$\%$  for $x$ =  0.023 and to $-$25 MHz/$\%$ for $x$ = 0.028, and $\nu_{Q0}$ of $\sim$ 26 MHz  is nearly independent of $x$. 
        Such linear relationship has also been reported in Co pnictides by Majumder ${\it et~al.}$ (Ref. \onlinecite{Majumder2013}) and  can be found in other itinerant magnetic systems such as the BaFe$_2$As$_2$ (Ref. \onlinecite{KitagawaBaFeAs}) and LiFeAs (Ref. \onlinecite{BeakLiFeAs}). 
      The  $\alpha$ values estimated from the slopes of the $\nu_Q$ vs. $K$ plot are ranged in $\sim$ 0.04 MHz/$\%$ for PrCoAsO, $\sim$ 4 MHz/$\%$  in LiFeAs, much less than that in Ca(Fe$_{1-x}$Co$_x$)$_2$As$_2$.  
       According to self-consistent renormalization (SCR) theory,\cite{Takahashi1978} temperature dependence of $\nu_Q$ can be influenced by the spin susceptibility due to the mode mode coupling between charge and spin density fluctuations. 
       Thus the prominent $\alpha$ value indicates the strong coupling between charge  and spin density fluctuations in Ca(Fe$_{1-x}$Co$_x$)$_2$As$_2$, consistent with the NMR spectrum data showing the strong coupling between lattice and magnetism.

\subsection{$^{75}$As spin lattice relaxation rates 1/$T_1$}

   In order to investigate the evolution of the spin dynamics with Co substitution, we have measured  $^{75}$As spin lattice relaxation rates 1/$T_1$ as a function of temperature.
    Figures \ \ref{fig:T1T}(a) and \ \ref{fig:T1T}(b) show 1/$T_1T$ versus $T$ in Ca(Fe$_{1-x}$Co$_{x}$)$_2$As$_2$ for $H$ perpendicular and parallel  to the $c$ axis at $H$ $\sim$ 7.5 T, respectively.  
        For $x$ = 0.023,  above $T_{\rm N}$ = 106 K,  1/$T_{1}T$  for $H$ $\parallel$ $ab$ plane shows a monotonic increase with decreasing $T$, while  1/$T_{\rm 1}T$ for $H$ $\parallel$ $c$ axis is nearly independent of $T$, similar to the case for $x$ = 0 reported previously.\cite{Furukawa2014} 
        At $x$ = 0.028, 1/$T_1T$ for both $H$ directions shows the similar behavior with those in $x$ $<$ 0.023 above $\sim$ 80 K, but 1/$T_1T$ starts to decrease below that temperature down to $T_{\rm N}$  = 53 K, suggesting a suppression of low energy spin excitations below $\sim$ 80 K. 
      In the case of superconducting samples with $x$ $\geq$ 0.033, 1/$T_{\rm 1}T$ for both magnetic field directions is nearly constant above $\sim$ 100 K but a suppression of  the spin excitations can be observed below $\sim$ 100 K.  
       With a further decrease of $T$,  1/$T_{1}T$ for $x$ = 0.033 and 0.059 shows a sudden decrease below $T_{\rm c}$ [15 (10) K for $x = $ 0.033 (0.059)] due to superconducting transitions, demonstrating not filamentary but bulk superconductivity in the system.
      This is consistent with the observation of clear jump at $T_{\rm C}$ in specific heat measurements.\cite{Ran2012}

\begin{figure*}[tb]
\includegraphics[width=14.0cm]{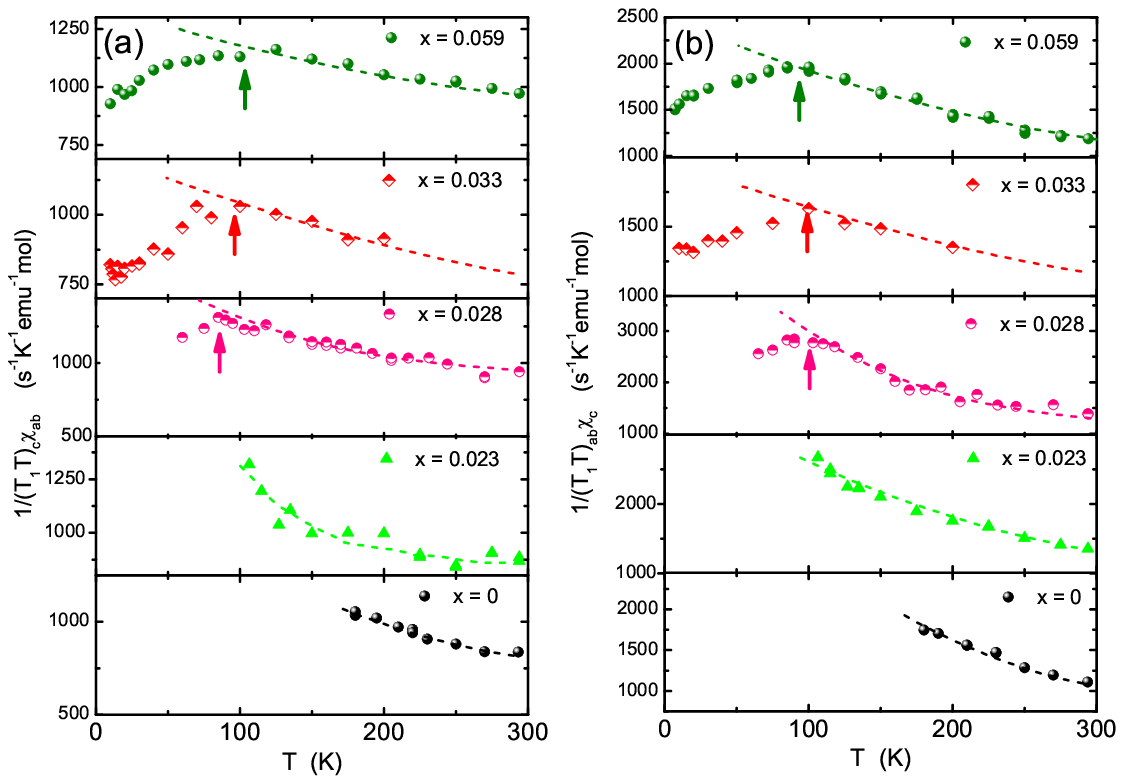} 
\caption{(Color online)  1/$T_1T$$\chi$  versus $T$ in the paramagnetic state for both magnetic field directions, 
(a) $H$ $\parallel$ $c$-axis and (b) $H$ $\parallel$ $ab$-plane. 
      The arrows indicate $T^*$ as discussed in the text. 
   The increases of  1/$T_1T$$\chi$ observed above $T^*$ indicates the growth of the stripe-type AFM spin correlations, while the decreases below $T^*$ indicate the suppression of the AFM spin correlations.    
   The dashed lines are guides for eyes.  }
\label{fig:T1Tchi}
\end{figure*}

\begin{figure}[tb]
\includegraphics[width=7.5cm]{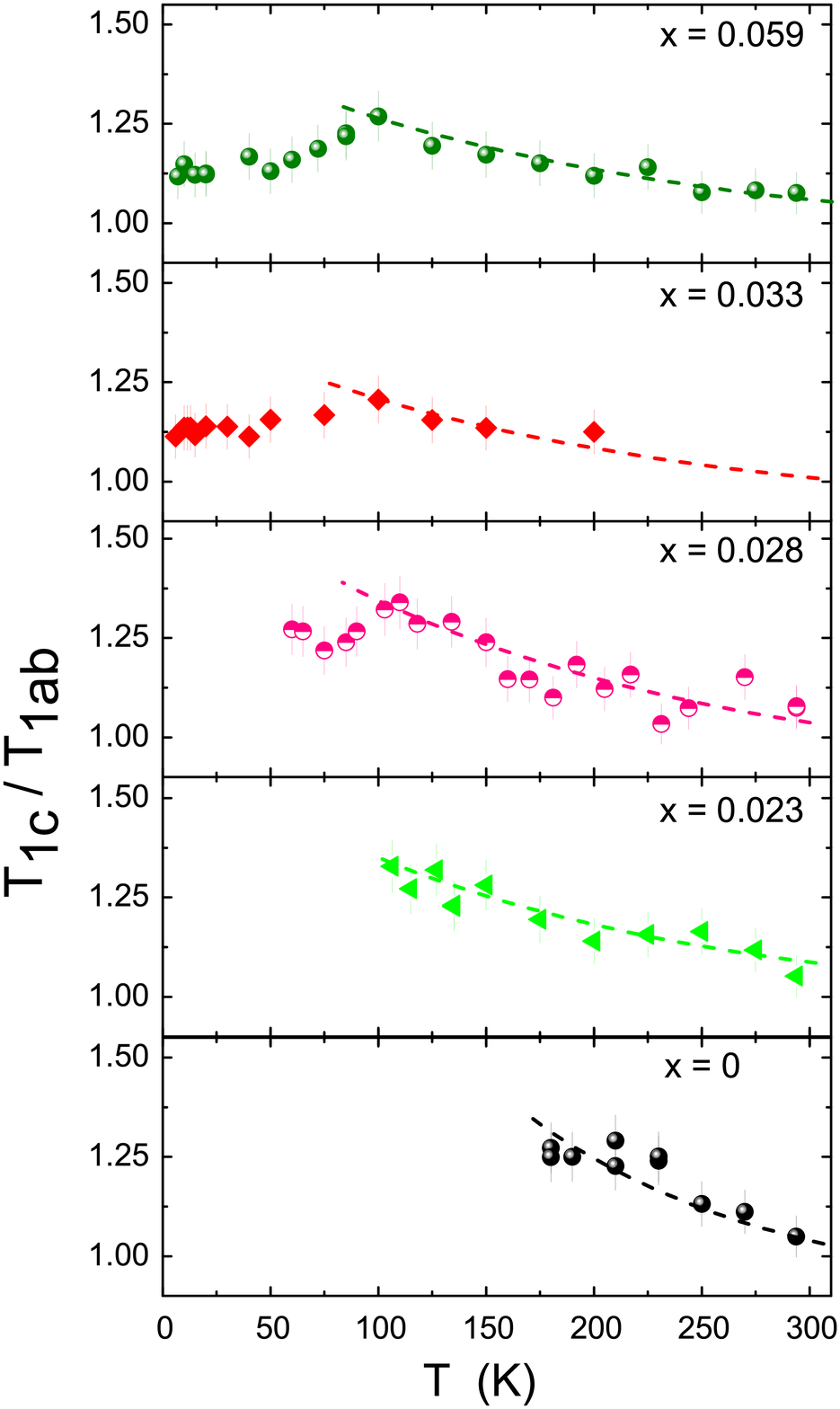} 
\caption{(Color online) $T$ dependence of the ratio $r$ $\equiv$ $T_{\rm 1,c}$/$T_{\rm 1,ab}$.   
   The dashed lines are guides for eyes. }
\label{fig:T1ratio}
\end{figure}

      In order to see AFM spin fluctuation effects in the paramagnetic state, it is useful to re-plot the data by changing the vertical axis from $1/T_1T$ to $1/T_1T\chi$ as shown in Fig.~\ref{fig:T1Tchi}, where the corresponding $\chi$ was used for each $H$ direction.\cite{Ran2012} 
    $1/T_{\rm 1}T$ can be expressed in terms of the imaginary part of the dynamic susceptibility $\chi^{\prime\prime}(\vec{q}, \omega_0)$ per mole of electronic spins as,\cite{Johnston2010,Moriya1963}
$\frac{1}{T_1T}=\frac{2\gamma^{2}_{N}k_{\rm B}}{N_{\rm A}}\sum_{\vec{q}}|A(\vec{q})|^2\frac{\chi^{\prime\prime}(\vec{q}, \omega_0)}{\omega_0}$, 
where the sum is over the wave vectors $\vec{q}$ within the first Brillouin zone, $A(\vec{q})$ is the form factor of the hyperfine interactions and $\chi^{\prime\prime}(\vec{q}, \omega_0)$  is the imaginary part of the dynamic susceptibility at the Larmor frequency $\omega_0$.  
    On the other hand,  the uniform $\chi$ corresponds to the real component 
 $\chi^{\prime}(\vec{q}, \omega_0)$ with $q$ = 0 and $\omega_0$ = 0. 
    Thus a plot of $1/T_{\rm 1}T\chi$ versus $T$ shows the $T$ dependence of  $\sum_{\vec{q}}|A(\vec{q})|^2\chi^{\prime\prime}(\vec{q}, \omega_0)$ with respect to that of the uniform susceptibility $\chi^{\prime}$(0, 0). 
    In order to eliminate effects of impurity contributions in the magnetic susceptibilities in our analysis, we used the magnetic susceptibility data which we subtracted impurity contributions from the original $\chi$ data by using our Knight shift data as has been done in  CaFe$_2$As$_2$ (Ref. \onlinecite{Furukawa2014}). 
    For above $T_{\rm N}$, $1/T_{\rm 1}T\chi$ for $H$ $\parallel$ $c$ axis and $H$ $\parallel$ $ab$ plane in all five samples increase with decreasing temperature. 
    The increase implies $\sum_{\vec{q}}|A(\vec{q})|^2\chi^{\prime\prime}(\vec{q}, \omega_0)$ increases  more than $\chi^{\prime}$(0, 0), which is due to a growth of spin correlations with $q$ $\neq$ 0 stripe-type AFM wave vector $q$ = $Q_{\rm AF}$,  as has been discussed in the $x$ = 0 case.\cite{Furukawa2014, Goldman2008}
     However, $x$ $\geq$ 0.028, one can clearly see that 1/($T_1T\chi$) are suppressed below a $T^*$-value marked by arrows. 
     We attribute this behavior to a pseudogap phenomenon.    
   The $T^*$s are nearly independent of Co substitution and are plotted in the phase diagram (Fig. 12). 
    It is noted that the our definition of the pseudogap behavior is corresponding to suppressions of the spin fluctuations with only the stripe-type AFM wave vectors not from $q$ = 0 components.
   The suppressions in spin fluctuations with the $q$ = 0 component can be seen in the temperature dependence of $K_{\rm spin}$ with nearly $x$-independent $\Delta$/$k_{\rm B}$ $\sim$ 490 K.

\begin{figure}[tb]
\includegraphics[width=8.5cm]{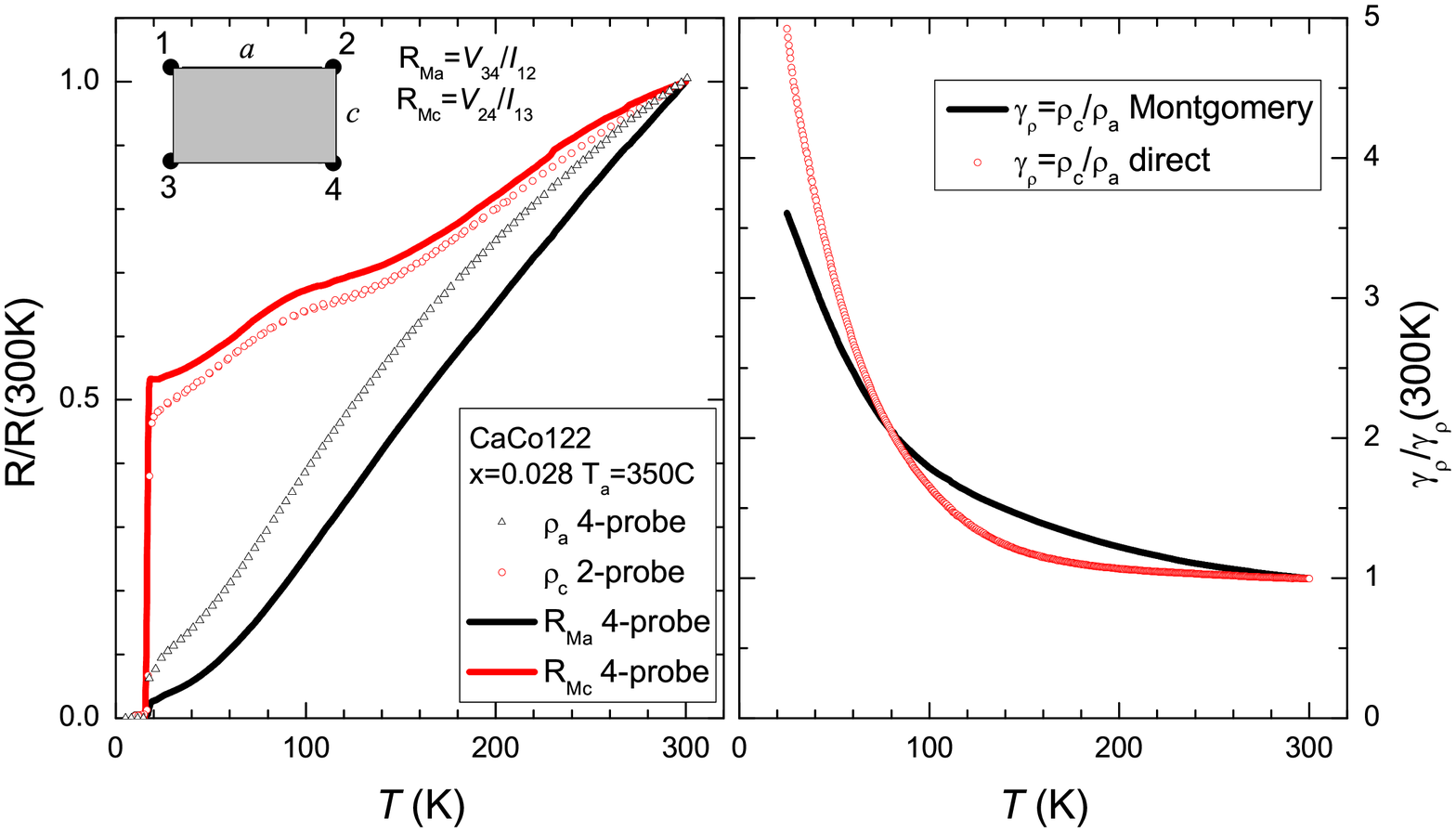} 
\caption{ (Color online) Temperature dependence of in-plane resistivity measured in CaCo122 samples of $x$=0.028, $T_{\rm a}$ = 350 $^{\circ}$C in four-probe configuration ($\rho_{\rm a}$, black triangles) and out-of-plane resistivity measured in two-probe configuration with contacts covering the whole $ab$-plane area ($\rho_{\rm c}$, red open circles). For comparison we show temperature-dependent resistances measured in four-probe Montgomery configuration with contacts located at the corners of the sample as shown schematically in the left panel. $R_{\rm Ma}=V_{34}/I_{12}$ was measured with current (flowing between contacts 1 and 2) and potential difference (between contacts 3 and 4) along the $a$-axis in the plane, and $R_{\rm Mc}=V_{44}/I_{13}$ with current and potential drop along the $c$ axis. All data are shown using normalized resistivity scale $R/R(300K)$. Raw Montgomery measurements represent weighted mixture of $\rho_a$ and $\rho_c$ with dominant contributions from respective current direction components. Comparison of $\rho_c(T)$ and $R_{MC}(T)$ directly shows that the features in the temperature dependent inter-plane resistivity are not affected by contacts covering the whole surface area of the sample. Right panel shows anisotropy ratio $\gamma_{\rho}\equiv \rho_c/\rho_a$, normalized to a room temperature value  $\gamma_{\rho} \approx$ 4, as determined from comparison of the direct resistivity measurements on two different samples in four- and two-probe configurations,  and measurements taken in Montgomery configuration on the same sample. The data are truncated at 25~K due to noise appearing from partial contribution of superconductivity. 
}
\label{fig:FigMontgomery}
\end{figure}

\begin{figure}[tb]
\includegraphics[width=7.5cm]{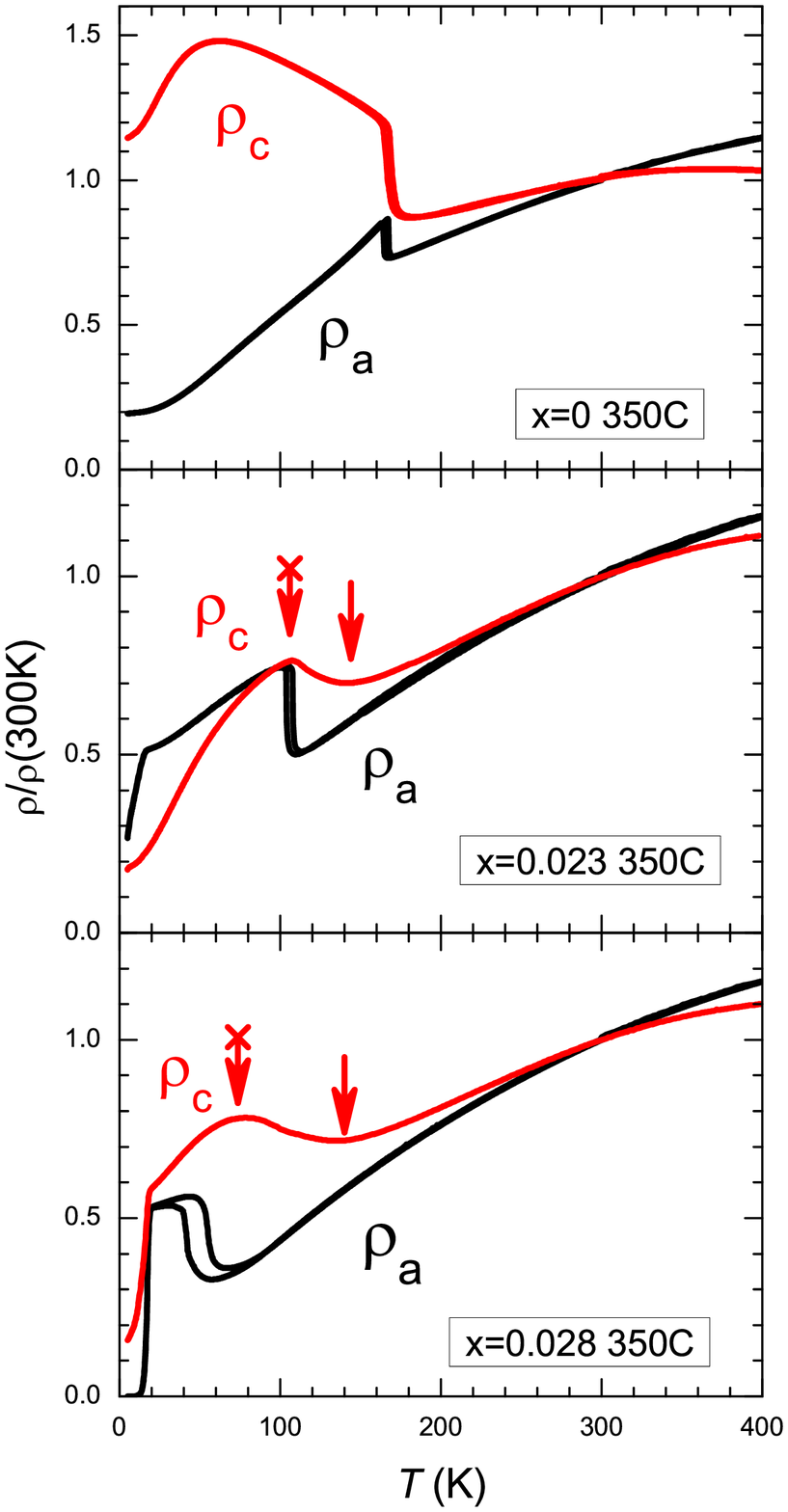} 
\includegraphics[width=7.5cm]{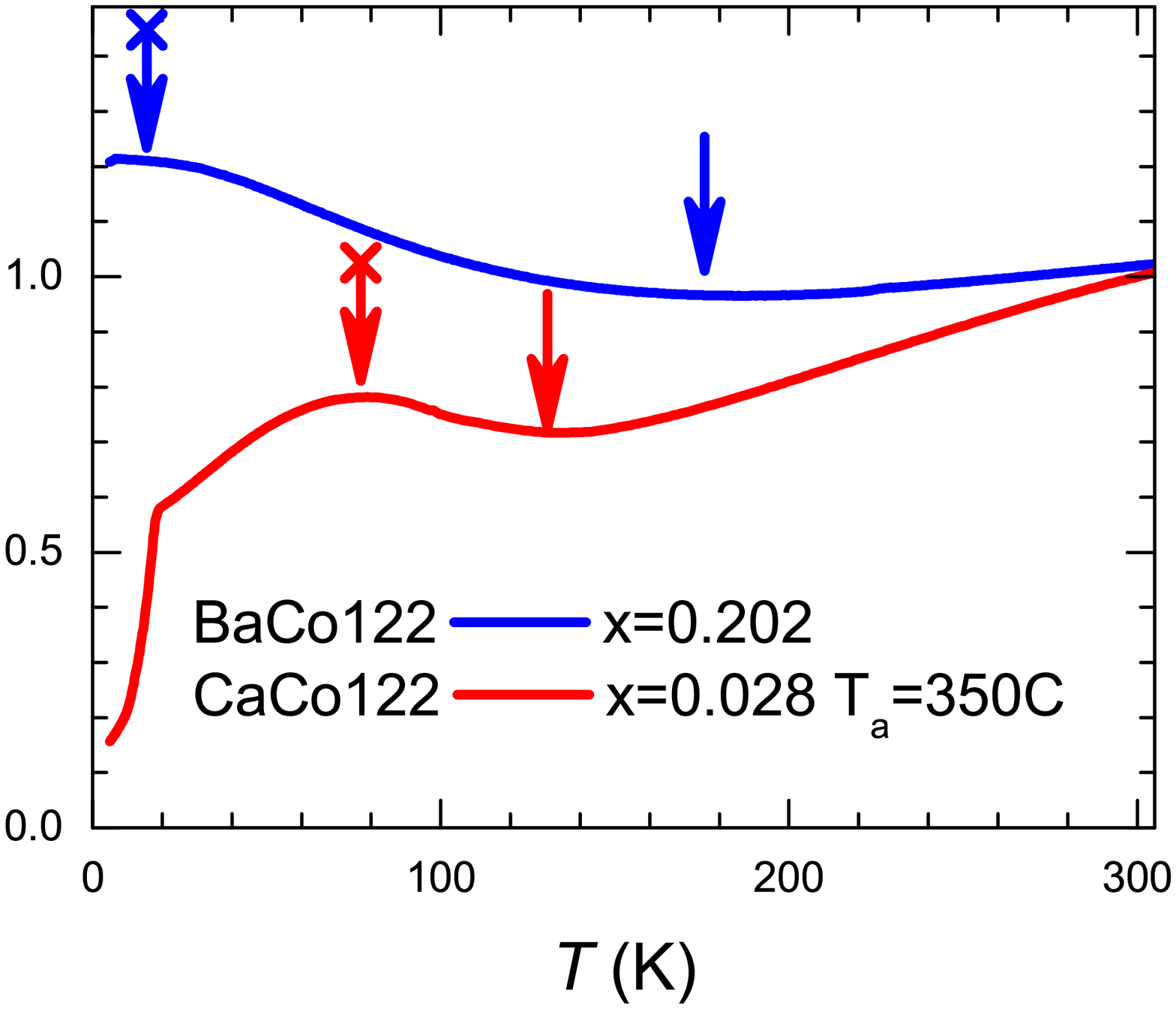} 
\caption{ (Color online) Temperature dependence of in-plane ($\rho_a$) and out-of-plane ($\rho_{c}$) resistivities for (top to bottom) $x$ = 0 ($T_{\rm a}$ = 400 $^{\circ}$C), 0.023 ($T_{\rm a}$ = 350 $^{\circ}$C), and 0.028 ($T_{\rm a}$ = 350 $^{\circ}$C),  (same batches as used in NMR measurements). 
The inter-plane resistivity, $\rho_c(T)$, shows broad minimum, denoted by straight arrow, and maximum, shown with cross-arrow. 
    In the bottom panel we compare $\rho_c(T)$ for sample $x$ = 0.028 to that of the sample of Ba(Fe$_{1-x}$Co$_x$)$_2$As$_2$, $x$ = 0.202 (non-superconducting heavily over-doped composition), showing similar features.\cite{Tanatar2010}
}
\label{fig:resistivity}
\end{figure}

\begin{figure}[tb]
\includegraphics[width=7.5cm]{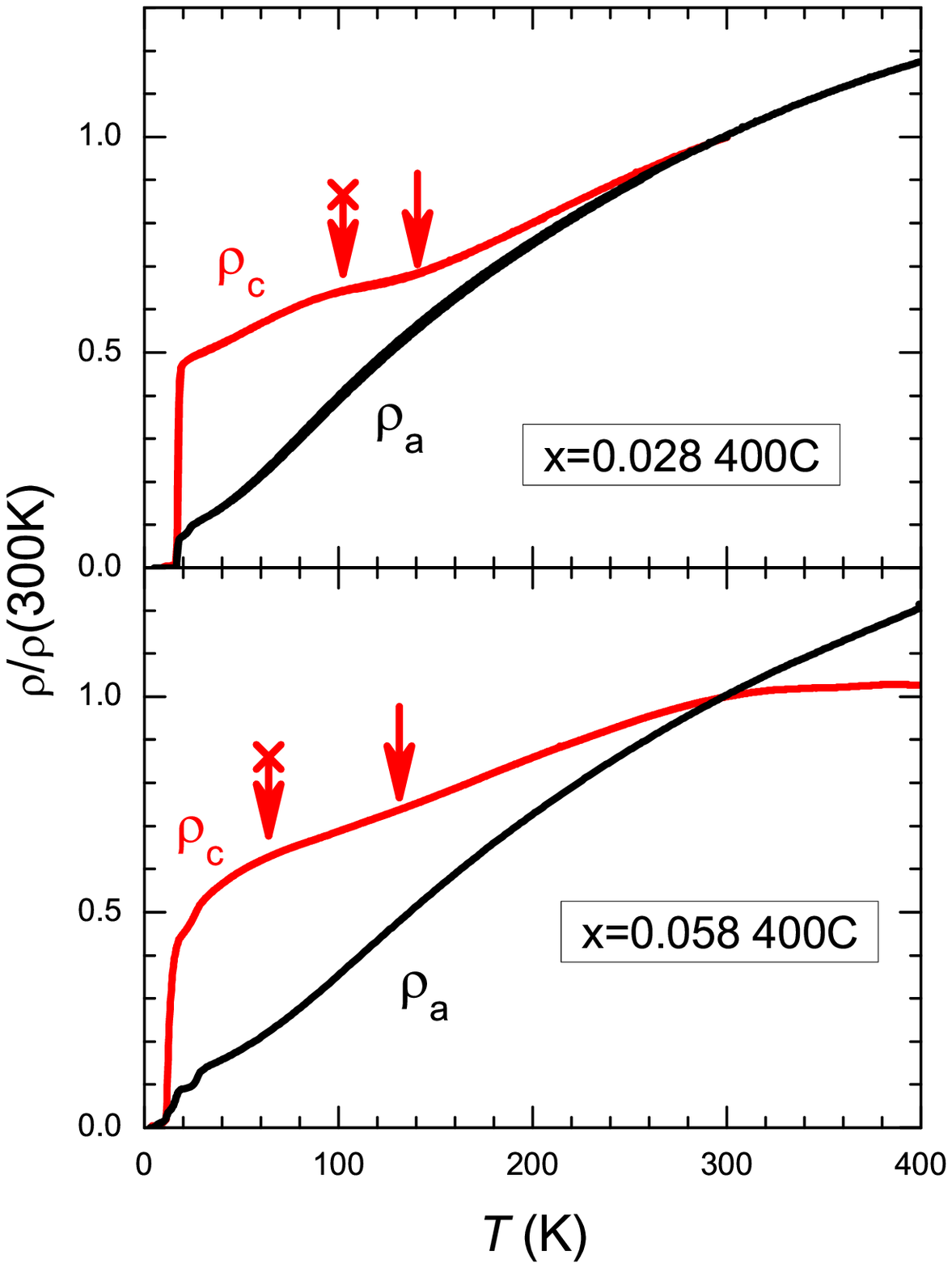} 
\includegraphics[width=7.5cm]{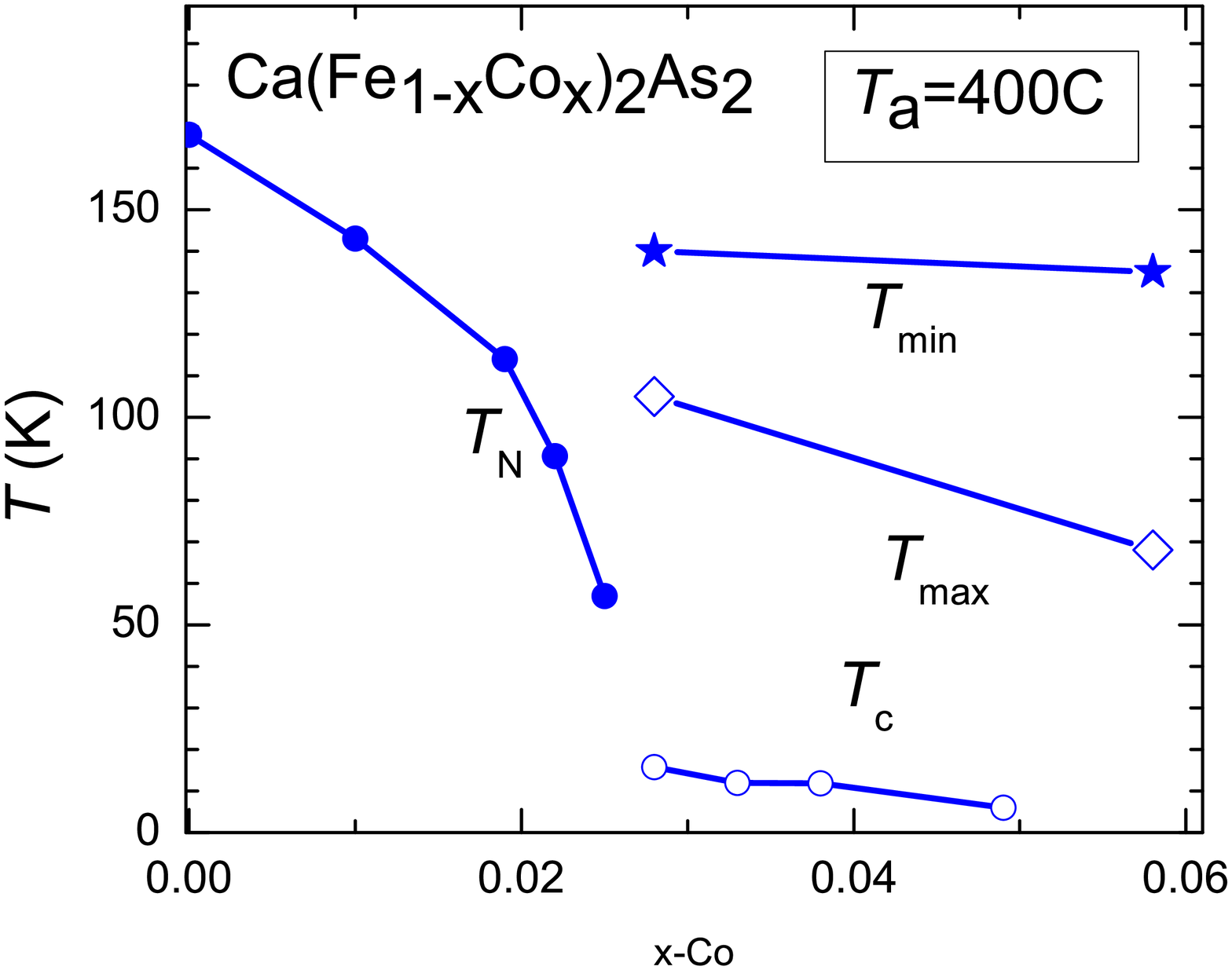} 
\caption{(Color online) Temperature dependence of in-plane ($\rho_a$) and out-of-plane ($\rho_{c}$) resistivities for samples with $x$ = 0.028 and 0.058 annealed at $T_{\rm a}$ = 400 $^{\circ}$C. 
    The samples are representative of superconducting and heavily over-doped non-superconducting regions. The $\rho_c(T)$ still shows minimum-maximum structure on cooling, suggesting the presence of pseudogap features through all the phase diagram. 
    Bottom panel shows the phase diagram as determined from resistivity measurements on samples with 400 $^{\circ}$C annealing. $T_{\rm N}$ and $T_{\rm c}$ for the samples with 400 $^{\circ}$C annealing are from Ref. \onlinecite{Ran2012}
}
\label{fig:resistivity400C}
\end{figure}

   Based on these $T_1$ results, we can discuss more details of the Fe spin fluctuations in the pseudogap-like phase. 
   According to previous NMR studies performed on Fe pnictides,\cite{KitagawaSrFe2As2, KitagawaAFSF, FukazawaKBaFe2As2} and SrCo$_2$As$_2$,\cite{PandeySrCo2As2}  
   the  ratio $r$ $\equiv$ $T_{1,c}$/$T_{1, ab}$ depends on AFM spin correlation modes as 
\begin{eqnarray}
r =  \left\{
\begin {array}{ll}

0.5 + \left(\frac{S_{ab}}{S_c}\right) ^2    \mbox{~ for  the stripe AFM fluctuations} \\ 
% 0.5 + \left(\frac{S_{ab}}{S_c}\right) ^2    \mbox{~ for}  ~ {\bf q} = (\pi,0)  \mbox{~or} ~ (0, \pi) \label{eqn:stripecorrelation}\\

0.5     \mbox{~ for  the N\'eel-type spin fluctuations} \\
% 0.5     \mbox{~ for}   ~ {\bf q} = (\pi,\pi) \label{eqn:Neelcorrelation}\\
\end {array}
\right .
\label{eqn:correlation}
\end{eqnarray}
where  ${\cal S}_{\alpha}$ is the amplitude of the spin fluctuation spectral density at NMR frequency along the $\alpha$ direction. 
    As plotted in Fig.\ \ref{fig:T1ratio}, the $r$ is greater than unity and, with decreasing $T$, $r$ increases up to $\sim$ 1.4 but never exceeds 1.5 even near $T_{\rm N}$ for $x$ $<$ 0.028.  
    This means that the stripe-type AFM fluctuations along the $c$ axis, ${\cal S}_{c}$, are stronger than the fluctuations in the $ab$ plane, ${\cal S}_{ab}$, in the paramagnetic phase, although ${\cal S}_{ab}$ are more enhanced than ${\cal S}_{c}$ with decreasing temperature. 
     An anisotropy in stripe-type AFM spin fluctuations is also observed in various Fe-based superconductors in the paramagnetic state\cite{FukazawaKBaFe2As2,Li2011} and the $r$ greater than 1.5  near $T_{\rm N}$ is observed in such as SrFe$_2$As$_2$ (Ref. \onlinecite{KitagawaSrFe2As2}) and LaFeAs(O$_{1-x}$F$_x$) (Ref. \onlinecite{Nakai2012}).  
   This indicates that ${\cal S}_{ab}$ is greater than ${\cal S}_{c}$ near $T_{\rm N}$, in contrast to our results for Ca(Fe$_{1-x}$Co$_x$)$_2$As$_2$.
    In the case of $x$ $\geq$ 0.028,  the $r$ increases with decreasing temperature as in the case of $x$ $<$ 0.023, but one can see clear decrease in $r$ below $\sim$ $T^*$ which is due to the presence of the pseudogap-like phase: suppressions of the stripe-type AFM spin fluctuations.  
     Interestingly, from the temperature dependence of $r$, the ${\cal S}_{ab}$ is found to be suppressed more than ${\cal S}_{c}$ in the pseudogap phase below $T^*$. 
     
    Finally it is interesting to discuss 1/$T_1T$ data  in the AFM state below $T_{\rm N}$ and compare with those in Ba(Fe$_{2-x}$Co$_x$)$_2$As$_2$.   
     As shown in Figs. 6(a) and 6(b), $1/T_1T$ suddenly dropped just below  $T_{\rm N}$ in Ca(Fe$_{1-x}$Co$_{x}$)$_2$As$_2$. 
   This originates from a sudden suppression of AFM spin fluctuations in the AFM state and a reduction of the density of states ${\cal D}(E_{\rm F})$ at the Fermi energy  due to a reconstruction of Fermi surface below $T_{\rm S}$.\cite{KitagawaBaFeAs,Ning2014} 
    The temperature dependence of 1/$T_1T$ is quite different from that of 1/$T_1T$ in Ba(Fe$_{1-x}$Co$_{x}$)$_2$As$_2$. 
   In Fig. 6(c), we compare  the 1/$T_1T$ data for the $x$ = 0.028 Co substituted CaFe$_2$As$_2$  with those for the $x$ = 0.047 in Ba(Fe$_{1-x}$Co$_{x}$)$_2$As$_2$ ($T_{\rm N}$  = 50 K, $T_{\rm C}$  = 15 K).\cite{Wiecki2015} 
  Here we chose the 4.7$\%$ Co substituted BaFe$_2$As$_2$ sample because $T_{\rm N}$  = 50  K is close to $T_{\rm N}$ = 53 K of the 2.8$\%$ Co substituted  CaFe$_2$As$_2$. 
    As seen in the figure, 1/$T_1T$ for the 4.7$\%$ Co substituted BaFe$_2$As$_2$ shows clear divergent behavior around $T_{\rm N}$ = 50 K and a gradual decrease below $T_{\rm N}$, contrast to the case of  the Co substituted  CaFe$_2$As$_2$.
  With further decrease in temperature, 1/$T_1T$ exhibits constant behavior at low temperatures with a sudden decrease below  $T_{\rm C}$  = 15 K due to SC transition, evidencing the coexistence of AFM and SC states. 
     In Fig. 6(d), the 1/$T_1T$ data for the 2.3$\%$ Co substituted CaFe$_2$As$_2$ are also compared with those for $x$ = 0.02 (non-SC) in Ba(Fe$_{1-x}$Co$_{x}$)$_2$As$_2$ (Ref. \onlinecite{Ning2014}) with $T_{\rm N}$ = 99 K close to $T_{\rm N}$ = 106 K of the 2.3$\%$ Co substituted  CaFe$_2$As$_2$.  
   We again see similar differences in the temperature dependence of 1/$T_1T$ between the samples.        
   Here one interesting feature in Ba(Fe$_{1-x}$Co$_{x}$)$_2$As$_2$ is that the reduction in 1/$T_1T$ below $T_{\rm N}$ is much less in the 4.7$\%$ Co substituted SC BaFe$_2$As$_2$ than in the 2$\%$ Co substituted non-SC BaFe$_2$As$_2$. 
   Although the 1/$T_1T$ values below $T_{\rm N}$ are related to the  ${\cal D}(E_{\rm F})$ and AFM spin fluctuations, the large value of 1/$T_1T$ at low temperatures,  comparable to the 1/$T_1T$ value at high temperatures, for the 4.7$\%$ Co substituted sample cannot be attributed to the change in the ${\cal D}(E_{\rm F})$ alone and may indicate that the significant AFM spin fluctuations still remain in the AFM state. 
   This would be consistent with the general idea that the AFM spin fluctuations play an important role for an occurrence of SC. 
    In addition to the different temperature dependence of 1/$T_1T$ below $T_{\rm N}$ between the  Co substituted CaFe$_2$As$_2$ and BaFe$_2$As$_2$ systems, the clear suppression of AFM spin fluctuations in the paramagnetic states below $T^*$ for $x$ $\geq$  0.028 in Ca(Fe$_{1-x}$Co$_{x}$)$_2$As$_2$ is not observed in Ba(Fe$_{1-x}$Co$_{x}$)$_2$As$_2$. \cite{Ning2009,Ning2014,Wiecki2015} 
   It is also noticed that the ordered Fe moments are robust with Co substitution in Ca(Fe$_{1-x}$Co$_{x}$)$_2$As$_2$, contrast to the case in Ba(Fe$_{1-x}$Co$_{x}$)$_2$As$_2$.\cite{Fernandes2010} 
   Therefore, it is likely that these differences lead to these Co substituted CaFe$_2$As$_2$ samples having no coexistence of AFM and SC.   
   
%   Finally it is worth to mention that the 1/$T_1T$ decreases below $T_{\rm N}$ milder in  the 4.7$\%$ Co substituted SC BaFe$_2$As$_2$ than in the 2$\%$ Co substituted non-Sc BaFe$_2$As$_2$. 
 %  Since the 1/$T_1T$ values below $T_{\rm N}$ are mainly determined by not only  ${\cal D}(E_{\rm F})$ and AFM spin fluctuations, this could be also consistent with the general idea that the AFM spin fluctuations play an important role for an occurrence of SC. 

\subsection{Temperature-dependent anisotropic resistivity}

In Fig.~\ref{fig:FigMontgomery}, we study how the measurements in the two-probe configuration, revealing the most clear signatures of the pseudogap, as we discuss below, are affected by the two-probe technique measurements. For this purpose we compare measurements taken on samples of CaCo122 $x$ = 0.028, $T_{\rm a}$=350 $^{\circ}$C, taken in standard four-probe contact configuration, $\rho_a(T)$, in two-probe configuration, $\rho_c(T)$ and in Montgomery technique measurements. In Montgomery technique, resistivity measurements are performed in four-probe contact scheme, as shown schematically in the left panel of Fig.~\ref{fig:FigMontgomery}. Two measurements are taken with current and potential drop along principal directions of conductivity tensor, $a$ and $c$ crystallographic directions. In the first measurement current is flowing between contacts 1 and 2, and potential drop is measured between contacts 3 and 4. 
   Thus determined values are used to calculate resistivity $R_{\rm MA}=V_{34}/I_{12}$, which is a weighted mixture of $\rho_a$ and $\rho_c$, with dominant contribution of $\rho_a$ depending on sample dimensions. Similar measurements along $c$-axis define resistance $R_{\rm MC}=V_{24}/I_{13}$ with dominant contribution of $\rho_c$. Direct comparison of measured $\rho_a$ and $R_{\rm Ma}$ and $\rho_c$ and $R_{\rm Mc}$ in Fig.~\ref{fig:FigMontgomery}, clearly shows that the features in $\rho_c(T)$ are observed at the same temperatures in both samples with full and partial coverage of the $ab$-plane with solder, thus showing that the effect of the contact stress are negligible. In the right panel we compare anisotropies determined and calculated from measurements on two different samples ($\rho_a$ and $\rho_c$) and calculated in Montgomery technique measurements of the same sample. In both cases we obtain $\gamma_{\rho}=\rho_c/\rho_a \approx $ 4 at 300~K and very similar temperature dependence with mild $\approx$ 4 times increase of anisotropy on cooling, clearly showing self-consistency of direct and Montgomery technique measurements of $\rho_c(T)$.

    In the top three panels of Fig.~\ref{fig:resistivity} we plot temperature-dependent in-plane and inter-plane resistivity, using normalized value $\rho(T)/\rho(300K)$. These measurements were performed on samples from the same batches as used in NMR study.
     Both $\rho_a(T)$ and $\rho_c(T)$ show an initial metallic decrease on cooling at temperatures above the sharp, hysteretic, jump signaling first order structural-magnetic transition at $T_{\rm N}$. 
   In the parent $x$ = 0 compound, the $\rho_c(T)$ starts to increase above $T_{\rm N}$, and this increase of resistivity on cooling continues down to approximately 50~K, signaling opening of the partial gap on the Fermi surface. 
    For $x$ $>$ 0,  the increase of $\rho_c$ starts significantly above $T_{\rm N}$ and is gradual. 
    The sharp hysteretic feature, observed in $\rho_a(T)$ at $T_{\rm N}$, is smeared in $\rho_c(T)$ and instead gradual decrease of $\rho_c(T)$ is observed at low temperatures. 
    The overall behavior of $\rho _c(T)$ in the $x$ = 0.023 and $x$ = 0.028 samples is strongly reminiscent of the dependence found in heavily over-doped BaFe$_2$As$_2$ compounds substituted with Co (Ref. \onlinecite{Tanatar2010}) and Rh (Ref. \onlinecite{pseudogap2}). 
   Direct comparison of inter-plane resistivities for over-doped BaCo122 $x=$ 0.202 and CaCo122 ($x$ = 0.028, $T_{\rm a}$ = 400 $^{\circ}$C) is shown in the bottom panel of Fig.~\ref{fig:resistivity}. 
    The two features on the overall metallic behavior of $\rho_c(T)$ correspond to opening of partial gap (resistivity minimum at $T_{\rm min}$) and the end of carrier activation over partial gap and restoration of metallic properties at lower temperatures (resistivity maximum at $T_{\rm max}$).\cite{Tanatar2010,pseudogap2}

    The minimum-maximum structure in the resistivity could be related to the stripe-type AFM spin fluctuations revealed by the NMR measurements. 
   Electron scattering in the normal state of iron-based superconductors is predominantly magnetic, \cite{Erick-NatureComm} and transformations of magnetic correlations with temperature are reflected in temperature dependent resistivity in two ways. 
    Since the stripe-type AFM spin fluctuations originate from interband correlations due to the multiband structure at the Fermi surface in the Fe pnictides, the opening of partial gap at the Fermi surface leading to the resistivity minimum may  suppress the stripe-type AFM spin fluctuations with $q$ = ($\pi$, 0) or (0, $\pi$) wavevectors as seen in the NMR measurements if the partial gap affects the interband correlations. 
   On the other hand, a decrease of scattering at the other wavevectors would lead to a decrease in resistivity on  further cooling (resistivity maximum). 

%   The appearance of temporal correlations at a ($\pi$,0) wavevector of stripe phase in resistivity measurements mimics opening of a gap and leads to a resistivity increase on cooling (resistivity minimum), while decrease of scattering at the other wavevectors leads to resistivity decrease on further cooling (resistivity maximum). 

     NMR data suggest that pseudogap features are observed even in superconducting and heavily over-doped compositions. 
   We were not able to find samples with $x >$ 0.028, suitable for inter-plane resistivity measurements, with identical to NMR measurements annealing, $T_{\rm a}=$ 350 $^{\circ}$C, conditions. 
   Therefore we studied samples with different annealing temperature $T_a=$ 400 $^{\circ}$C, with $x=$ 0.028 (bulk superconductivity region; $T_{\rm c}$ = 15.7 K) and $x$ = 0.058 (heavily over-doped region of non-bulk superconductivity). 
   The top panel in Fig.~\ref{fig:resistivity400C} shows temperature-dependent resistivity of these samples. For both compositions the minimum-maximum structure is preserved in $\rho_c(T)$, with no corresponding features in $\rho_a(T)$. 
      In the bottom panel of Fig.~\ref{fig:resistivity400C} we plot phase diagram as determined from resistivity measurements on samples with $T_{\rm a}=$ 400 $^{\circ}$C. 
    This diagram suggests that the pseudogap features detected by the inter-plane resistivity measurements are observed in all substitution range from parent under-doped to heavily over-doped compositions. 
   This is consistent with NMR data, though due to a broad cross-over character of the features and the ambiguity of the criteria for the definitions of the characteristic temperatures, there is no direct correspondence between the two. Additional source of discrepancy between two data sets can come from difference of characteristic time scale of the two measurements. Resistivity measurements see magnetic correlations on a time scale of scattering time (of order of 10$^{-12}$ sec), while NMR measurements probe correlations at a much longer time scale (of order of 10$^{-6}$ sec). It is natural then to expect that the appropriate features happen at somewhat lower temperatures in NMR measurements, which seems to be the case. 
   Further studies for detail relationship between the resistivity minimum-maximum structure and the stripe-type AFM spin fluctuations are of clear interest.

\begin{figure}[tb]
\includegraphics[width=8.5cm]{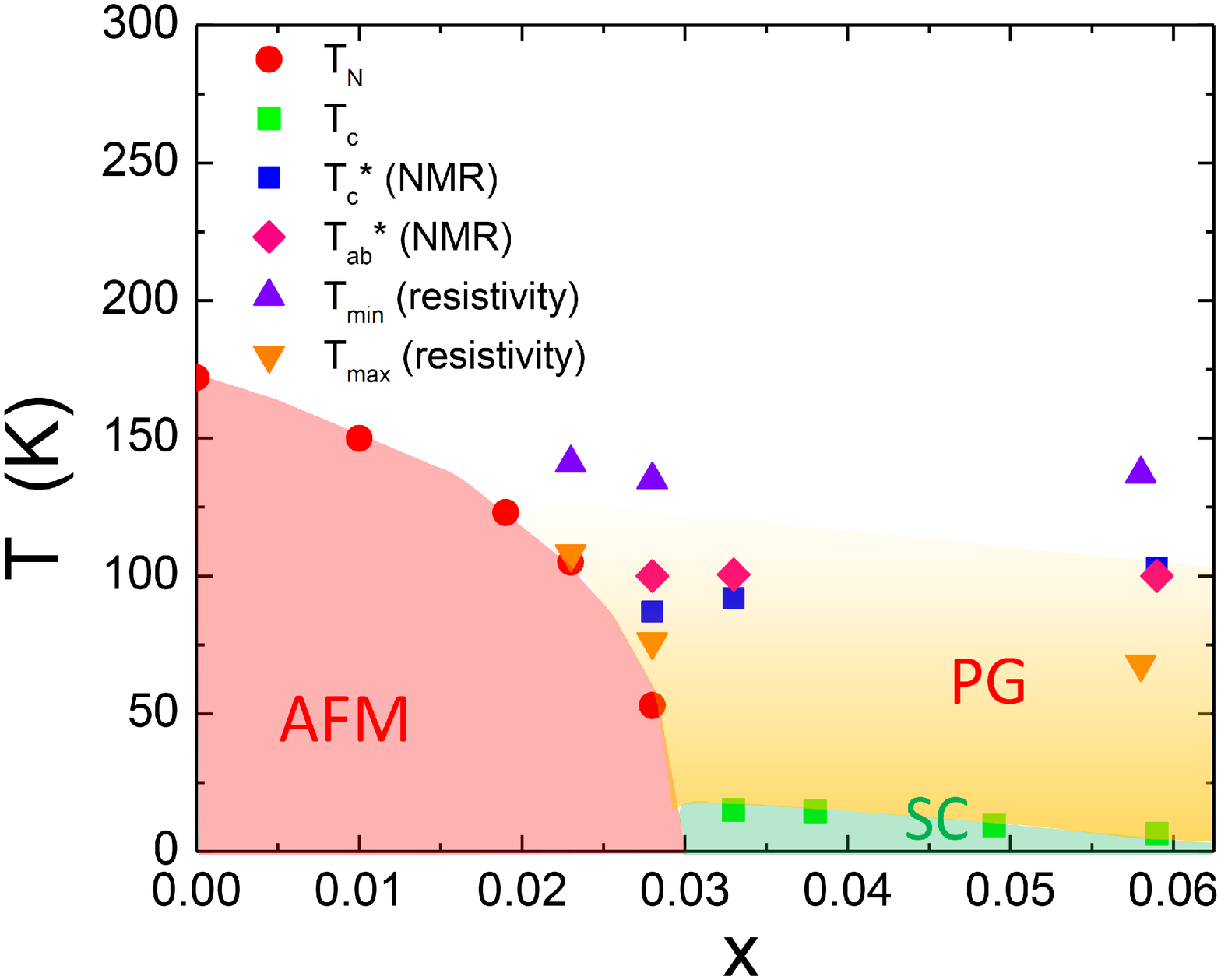} 
\caption{(Color online) Phase diagram of Ca(Fe$_{1-x}$Co$_x$)$_2$As$_2$. 
$T_{\rm N}$ and $T_{\rm c}$ are from Ref. \onlinecite{Ran2012}.
The crossover temperature $T_{ab}^*$ and $T_{c}^*$ are determined by NMR measurements for $H$  $\parallel$ $ab$ plane and  $H$  $\parallel$ $c$ axis, respectively. 
   $T_{\rm max}$ and $T_{\rm min}$ are estimated from the inter-plane resistivity measurements for the crystals annealed at $T_{\rm a}$ = 350 $^{\circ}$C except for $x$ = 0.058 with $T_{\rm a}$ = 400 $^{\circ}$C.      
AFM, SC and PG stand for the antiferromagnetic ordered state, superconducting, and pseudogap-like phases, respectively. 
}
\label{fig:As-spectrum}
\end{figure}

 \section{Summary} 
      Co substitution effects on static and dynamic magnetic properties of  the single-crystalline  Ca(Fe$_{1-x}$Co$_x$)$_2$As$_2$ ($x$  = 0, 0.023, 0.028, 0.033, 0.059) have been investigated by $^{75}$As nuclear magnetic resonance (NMR) and resistivity measurements.
      As in the case of $x$ = 0 ($T_{\rm N}$ = 170 K), clear evidence for the first order structural and stripe-type antiferromagnetic (AFM)  is observed from the sudden change in nuclear quadrupolar frequency ($\nu_{\rm Q}$) and internal field ($H_{\rm int}$)  at As sites  in  $x$ = 0.023 ($T_{\rm N}$ = 106 K) and $x$ = 0.028 ($T_{\rm N}$ = 53 K). 
      In the stripe-type AFM state, for magnetic field $H$ parallel to the $c$ axis, the observed clear separations of $^{75}$As NMR lines due to the internal field $H_{\rm int}$ indicate the commensurate stripe-type AFM state in the Co substituted crystals in $x$ = 0.023 ($T_{\rm N}$ = 106 K) and $x$ = 0.028 ($T_{\rm N}$ = 53 K), as in the case of $x$ = 0 ($T_{\rm N}$ = 170 K), similar to the case of Co/Ni substituted BaFe$_2$As$_2$. 
      Although $T_{\rm N}$ is strongly suppressed  from  170 K ($x$ = 0) to $x$ = 0.023 ($T_{\rm N}$ = 106 K) and $x$ = 0.028 ($T_{\rm N}$ = 53 K) with Co substitution, $H_{\rm int}$ decreases only slightly from 2.64 T to 2.35 T and 2.25 T respectively, suggesting robustness of the Fe magnetic moments upon Co substitution in Ca(Fe$_{1-x}$Co$_x$)$_2$As$_2$.
     In the paramagnetic state, the temperature dependence of Knight shift $K$ for all crystals shows similar temperature dependence of magnetic susceptibility, where the temperature dependent part of $K$ can be fitted with a thermal activation behavior of exp(--$\Delta$/$k_{\rm B}$$T$) with nearly $x$ independent $\Delta$/$k_{\rm B}$ $\sim$ 490 K.
      These results indicate that spin fluctuations with the $q$ = 0 components are suppressed with $\Delta$/$k_{\rm B}$ $\sim$ 490 K  in the paramagnetic state.

       On the other hand, the growth of the stripe-type AFM fluctuations with $q$ = ($\pi$, 0) or (0, $\pi$) on lowering temperature in the paramagnetic state for all five crystals is evidenced by the temperature dependence of the nuclear spin lattice relaxation rates divided by temperature and magnetic susceptibility (1/$T_1T\chi$). 
    In addition, above $x$ $\geq$  0.028,  1/$T_1$$T$$\chi$ is found to show a gradual decrease with decreasing $T$ below $T^*$, a crossover temperature, corresponding to suppression of the stripe-type AFM fluctuations; attributed to the behavior of pseudogap-like phenomenon. 
    As shown in Fig. 12, $T^*$ $\sim$ 100 K is almost independent of $x$.  
    It is pointed out that the pseudogap-like phenomenon seems to affect on the temperature-dependent inter-plane resistivity, $\rho_c(T)$, but not with in-plane resistivity $\rho _a (T)$. 
     The ratio of $1/T_1$ for magnetic fields $H$ parallel to the $ab$ plane and to the $c$ axis, that is, $r$ = $(T_1)_c/(T_1)_{ab}$, increases with decreasing $T$ and starts to decreases below $T^*$. 
     This indicates that the amplitude of stripe-type AFM fluctuations in the $ab$ plane (${\cal S}_{ab}$) is more enhanced than that along the $c$ axis (${\cal S}_{c}$) above $T^*$, but ${\cal S}_{ab}$ is more suppressed than ${\cal S}_{c}$ in the pseudogap-like phase. 
     Further detailed studies on the pseudogap-like phase in Ca(Fe$_{1-x}$Co$_x$)$_2$As$_2$ will be required to shed the light on, using other experimental techniques such as inelastic neutron scattering measurements having different energy scale from NMR technique.

\section{Acknowledgments}
The work was supported by the U.S. Department of Energy (DOE), Office of Basic Energy Sciences, Division of Materials Sciences and Engineering. The research was performed at Ames Laboratory, which is operated for the U.S. DOE by Iowa State University under Contract No.~DE-AC02-07CH11358.

\end{document}